\documentclass[preprint,12pt]{elsarticle0}

\usepackage{hyperref}
\usepackage{graphicx}
\usepackage{subfigure}
\usepackage{amssymb}
\usepackage{array,multirow}
\usepackage{amsmath}
\usepackage{mathrsfs}
\usepackage{fancyhdr}
\usepackage{verbatim}
\usepackage{color}
\usepackage{bbm}
\usepackage{epstopdf}

\newcommand{\MSb}{\overline{\mathrm{MS}}}

\newcommand{\oo}{\mathcal{O}}
\newcommand{\lat}{\mathrm{lat}}
\newcommand{\free}{\mathrm{free}}
\newcommand{\massless}{\mathrm{massless}}
\newcommand{\cont}{\mathrm{cont}}

\usepackage{amssymb}
\usepackage{pstricks}

\begin{document}
\begin{frontmatter}

\title{Non-perturbative running of renormalization constants from correlators 
in coordinate space using step scaling}

\author{Krzysztof Cichy}
\ead{kcichy@th.physik.uni-frankfurt.de}
\address{Goethe-Universit\"at Frankfurt am Main, Institut f\"ur Theoretische Physik,
Max-von-Laue-Strasse 1, D-60438 Frankfurt am Main, Germany}
\address{Adam Mickiewicz University, Faculty of Physics, Umultowska 85, 61-614 Pozna\'n, Poland}

\author{Karl Jansen}
\address{NIC, DESY, Zeuthen, Platanenallee 6, 15738 Zeuthen, Germany}
\ead{karl.jansen@desy.de}

\author{Piotr Korcyl}
\ead{piotr.korcyl@ur.de}
\address{Institut f\"ur Theoretische Physik, Universit\"at Regensburg, D-93040 Regensburg, Germany}

\date{3 October 2016}

\begin{abstract}
Working in a quenched setup with Wilson twisted mass valence fermions, we explore the possibility to compute non-perturbatively the step scaling function using the coordinate (X-space) renormalization scheme. This scheme has the advantage of being on-shell and gauge invariant.
The step scaling method allows us to calculate the running of the renormalization constants of quark bilinear operators.
We describe here the details of this calculation. 
The aim of this exploratory study is to identify the feasibility of the X-space scheme when used in small volume simulations required by the
step scaling technique. Eventually, we translate our final results to the continuum $\MSb$ scheme and compare against four-loop analytic formulae finding satisfactory agreement. 
\end{abstract}


\end{frontmatter}
\rput(12.3cm,17.3cm){DESY 16-158}

\section{Introduction}
In lattice Monte Carlo simulations of QCD, connecting the high energy regime where perturbation theory can be safely applied,
 to the low energy regime of large volume simulations where hadron matrix elements are evaluated, usually cannot be done using a single
ensemble. 
The problem to safely treat the two very seperated scales within a single lattice extent and lattice spacing
is usually called the `window' problem and is summarized by the following equation in momentum space
\begin{equation}
\Lambda^2_{\textrm{QCD}} \ll p^2 \ll \frac{\pi^2}{a^2},
\end{equation}
or equivalently as
\begin{equation}
a^2 \ll x^2 \ll \frac{1}{\Lambda^2_{\textrm{QCD}}}
\end{equation}
in position space.
The latter says that although the relevant physical distances at which we evaluate our observables should be sufficiently small,
we are limited by the discretization errors. At the same time, when we try to study larger distances where cut-off effects are expected
to be smaller, we may approach the intrinsic scale of QCD too closely and perturbation theory will not be applicable.
A method to connect those two regimes while keeping all systematic errors under control
was developed in Refs.~\cite{Luscher:1993gh,Luscher:1991wu,Jansen:1995ck,Bode:2001jv}. It is called step scaling and consists in performing
several simulations at a smaller and smaller
lattice spacing as well as smaller and smaller volumes starting at the coarse, large volume level and repeating the procedure
until the mentioned high energy regime of the theory can be reached.

The most popular implementation of the step scaling method is used to compute the running of the strong coupling constant and
uses Schr\"odinger functional (SF) boundary conditions \cite{Luscher:1992an}. Although it necessitates the implementation of boundary conditions which differ from the ones usually used in large volume simulations, such framework offers many advantages. 
By choosing apropriate field values at the time boundaries $t=0$ and $t=T$, one removes problematic zero modes of
the gauge field, hence significantly simplifying perturbative calculations. 
By the same token, the Dirac operator develops an energy gap which allows to simulate dynamical fermions at vanishing quark mass. 
The SF framework offers also additional freedom in constructing correlation functions, which can be used to implement renormalization or improvement conditions that have significantly reduced cut-off effects. 
Within the SF scheme, also the running of renormalization factors for the first moment of the quark non-singlet parton distribution has been computed \cite{Guagnelli:1998ve,Guagnelli:1999gu}.

An alternative implementation of the step scaling technique was described in Ref.~\cite{Arthur:2010ht}. It uses the RI-MOM renormalization 
scheme \cite{Martinelli:1997zc}, with twisted boundary conditions in order to enable an easy implementation of renormalization conditions 
at a fixed physical momentum on the set of ensembles. 
Hence, instead of tuning the parameters of the entire simulated volume to stay on the line of constant physics, one chooses the twist angles 
such that the particular momentum used to define the renormalization condition remains fixed. The results of this implementation were
used for phenomenology among others in Ref.~\cite{Blum:2014tka}.

In the current work, we investigate an alternative to the RI-MOM implementation of the step scaling technique. 
We test renormalization conditions imposed on correlation functions in coordinate space (X-space) following Refs.~\cite{Martinelli:1997zc,Becirevic:2002yv,Gimenez:2003rt,Gimenez:2004me,Cichy:2012is} at sufficiently small distances. 
The same renormalization scheme can be used in a large volume simulation, hence we avoid implementing different boundary conditions. The aim of our work is to
test the precision and the cost of this implementation of the step scaling technique. 

Due to the known difficulties of perturbation theory for QCD in a small periodic box (see for example Ref.~\cite{vanBaal:1988qm} 
or the recent discussion in Ref.~\cite{Fodor:2012td} in the context of gradient flow), we look at correlation functions at very short distances (of the order of 0.01-0.1 fm) compared to the physical extent of the simulated box (which is an order of magnitude larger). This allows us to use infinite volume perturbation theory to translate our results to the $\MSb$ scheme. 
This solution comes with a certain price, as we have to simulate effectively large lattices in lattice units. 

The remainder of the present paper is organized as follows. In Section \ref{sec. position space}, we recall the most important details of the X-space renormalization scheme and the characteristics of the twisted mass formulation of lattice QCD which is used as the discretization of valence quarks.
We describe the generation and tuning of the small volume ensembles.
We pay particular attention to the matching of physical volumes, which allows us to perform continuum extrapolations along the lines of constant
physics. 
Next, we consider possible sources of systematic errors in Section \ref{sec. systematic}. 
We discuss, in particular, uncertainties related to the mismatch of the tuned ensembles, discretization effects 
 and finite volume effects.
Finally, we present numerical results for the running of renormalization constants between energy scales of around 1.5 GeV and 17 GeV in all 
four channels (pseudoscalar, scalar, vector and axial vector) in Section \ref{sec. results}. Discussion of results and an outlook is presented in Section \ref{sec. conclusions}.

\section{Definitions}
\label{sec. position space}

\subsection{Lattice setup}
\label{sec. tmqcd}

This feasibility study is performed in the quenched approximation. 
This means that sea quarks are infinitely massive, i.e. effectively there are $N_f=0$ active quark flavours.
Therefore, the obtained results, extrapolated to the continuum limit, can be compared with continuum perturbation 
theory setting $N_f=0$.

\subsubsection{Gauge action}

We choose the Wilson plaquette gauge action,
\begin{equation}
 S_G[U] = \frac{\beta}{3}\sum_x\Big( \sum_{\substack{
      \mu,\nu=1\\1\leq\mu<\nu}}^4 \textrm{Re\,Tr} \big( 1 - P_{x;\mu,\nu}\big) \Big),
\end{equation}
with $\beta=6/g_0^2$, $g_0$ -- bare coupling, $P_{x;\mu,\nu}$ -- plaquette at spacetime point $x$ in the plane $\mu\nu$.

To generate the configurations, we use the CHROMA software \cite{Edwards:2004sx}, in its scalar or parallel version, 
depending on the lattice size. The chosen simulation algorithm is the standard heatbath with four overrelaxation steps.

\subsubsection{Valence quarks}
The computation of X-space correlators needs an introduction of valence quarks 
and for this, we use the twisted mass (TM) formulation of lattice QCD \cite{Frezzotti:2000nk,Frezzotti:2003ni,
Frezzotti:2004wz,Shindler:2007vp}, which is given in the so-called
twisted basis by
\begin{equation}
 S_l[\psi, \bar{\psi}, U] = a^4 \sum_x \bar{\chi}_l(x) \big( D_W + m_0 + i \mu_v \gamma_5 \tau_3
\big)
\chi_l(x),
 \label{tm_light}
\end{equation}
where $\tau^3$ is the third Pauli matrix acting in flavour space and $\chi_l=(\chi_u,\,\chi_d)$ is a
two-component vector in flavour space, related to the one in the physical basis by a chiral rotation.
$m_0$ and $\mu_v$ are the bare untwisted and twisted quark masses.
The massless Wilson-Dirac operator $D_W$ reads:
\begin{equation}
 D_W = \frac{1}{2} \big( \gamma_{\mu} (\nabla_{\mu} + \nabla^*_{\mu}) - a \nabla^*_{\mu} \nabla_{\mu}
\big),
\end{equation}
where $\nabla_{\mu}$ and $\nabla^*_{\mu}$ are the forward/backward covariant
derivatives.

One of the major advantages of TM fermions is that they are automatically $\mathcal{O}(a)$-improved by tuning just one parameter, the twist angle $\omega$, to maximal twist ($\omega=\pi/2$). This can be achieved by setting the hopping parameter $\kappa = (8+2
a m_0)^{-1}$ to its critical value, such that the PCAC quark mass vanishes
\cite{Frezzotti:2003ni,Chiarappa:2006ae,Farchioni:2004ma,Farchioni:2004fs,Frezzotti:2005gi,
Jansen:2005kk}.

\subsection{Coordinate space (X-space) renormalization scheme}

The X-space renormalization scheme was initially suggested in Ref.~\cite{Martinelli:1997zc} and first applied in 
Refs.~\cite{Becirevic:2002yv,Gimenez:2003rt,Gimenez:2004me} in the quenched approximation.
The first application in the dynamical case was reported by us in Ref.~\cite{Cichy:2012is}, where important improvements were implemented, and recently the method has been used, with further improvements, in Refs.~\cite{Tomii:2014lka,Tomii:2016xiv}.

Here, we shortly summarize the main ideas of this renormalization scheme.
For details, we refer to the above mentioned references.
We consider flavour non-singlet correlation functions of two operators of the form 
\begin{equation}
C_\Gamma(X)\equiv\langle \oo_{\Gamma}(X) \oo_{\Gamma}(0)\rangle, 
\end{equation}
where
\begin{equation}
 \oo_{\Gamma}(X) = \bar{\psi}(X) \Gamma \psi(X),\qquad \Gamma = 
 \{ 1,\gamma_5, \gamma_{\mu}, \gamma_{\mu} \gamma_5 \}\equiv\{S,P,V,A\}.
\end{equation}
We impose the following coordinate space conditions in the chiral limit,
\begin{equation}
 \lim_{a\rightarrow 0}\langle \oo^X_{\Gamma}(X) \oo^X_{\Gamma} (0) \rangle
\big|_{X^2=X_0^2} = \langle \oo_{\Gamma}(X_0) \oo_{\Gamma}(0)
\rangle^{\free, \massless}_{\cont},
\end{equation}
where we denote four-vectors with capital letters, e.g.~$X=(x,y,z,t)$ and $X^2\equiv x^2+y^2+z^2+t^2$.
The renormalized operator is
\begin{equation}
 \oo^X_{\Gamma}(X, X_0) = Z^X_{\Gamma}(X_0) \oo_{\Gamma}(X), 
\end{equation}
$X_0$ is the renormalization point, which we choose to satisfy 
\begin{equation}
 a\ll \sqrt{X_0^2} \ll \Lambda^{-1}
\end{equation}
in order to keep discretization effects under control and to ease contact to perturbation theory.
Here, $\Lambda$ denotes a low-energy scale of the considered theory and is of the order of a few hundred MeV.

It was shown in Ref.~\cite{Cichy:2012is} that the method works best for the so-called ``democratic'' points, which are defined
as points with a direction close to the diagonal of the hypercube (with an angle of at maximum 30 degrees between a given 
vector $X$ and $(1,1,1,1)$). 
However, further studies of this issue in the free theory suggest that certain points with an angle of 45 degrees are also expected to yield relatively small cut-off effects.

To alleviate the impact of cut-off effects, we use a tree-level correction, as defined in Ref.~\cite{Cichy:2012is}. This boils down to computing the ratio of the tree-level lattice and continuum correlators,
\begin{equation}
\label{eq:delta}
 \Delta_{\Gamma}(X) = \frac{\langle \oo_{\Gamma}(X) \oo_{\Gamma}(0)
\rangle_{\lat}^{\free}}{\langle \oo_{\Gamma}(X) \oo_{\Gamma}(0)
\rangle_{\cont}^{\free}}=
\frac{\langle \oo_{\Gamma}(X) \oo_{\Gamma}(0)
\rangle_{\lat}^{\free}}{\frac{c}{\pi^4(X^2)^3}},
\end{equation}
where $c=3$ for (pseudo)scalar correlators (later referred to also as PP/SS correlators), $c=6$ for (axial) vector ones (AA/VV correlators).
The corrected correlation function, $C'_{\Gamma}(X)$, reads
\begin{equation}
\label{eq:cprim}
 C'_{\Gamma}(X) = \frac{C_{\Gamma}(X)}{\Delta_{\Gamma}(X)}.
\end{equation}
The (tree-level corrected) renormalization constants, at the scale $\mu=1/\sqrt{X_0^2}$, are then given in the X-space scheme by
\begin{equation}
 Z^X_{\Gamma}(X_0) = \sqrt{\frac{C_{\Gamma}(X_0)_{\cont}^{\free}}{C'_{\Gamma}(X_0)}}
= \sqrt{\frac{C_{\Gamma}(X_0)_{\lat}^{\free}}{C_{\Gamma}(X_0)}}.
\end{equation}
In a certain case, we will also consider non-corrected renormalization constants,
\begin{equation}
 Z^X_{\Gamma}(X_0) = \sqrt{\frac{C_{\Gamma}(X_0)_{\cont}^{\free}}{C_{\Gamma}(X_0)}}
\end{equation}
and we will indicate it explicitly when this is the case (else we use the corrected ones).

One now needs to decide which points $X_0$ are best suited to extract the renormalization constants. We discuss
this issue in the next subsection.

\subsection{Step scaling}
In the step scaling method, one aims at estimating the step scaling function describing
the running of a scale dependent quantity as the renormalization scale is changed by some factor. In this work,
we fix this factor to 2. 
The finite lattice spacing version of the step scaling function can be constructed as a ratio of renormalization constants in the X-space scheme at 
two renormalization scales which differ by a given factor. 
Then, such ratio, which we denote by $\Sigma^X_{\Gamma}(\mu, 2\mu)$, 
has a well-defined continuum limit,
\begin{equation}
\Sigma^X_{\Gamma}(\mu, 2\mu) = \lim_{a \rightarrow 0} \frac{Z^X_{\Gamma}(2 \mu, a)}{Z^X_{\Gamma}(\mu, a)}, 
\label{eq. sigma}
\end{equation}
with $\mu=1/\sqrt{X_0^2}$ and where we have added explicitly the lattice spacing $a$ as an argument of $Z_\Gamma^X$ to indicate that it was regularized on the lattice.
The step scaling function enables one to non-perturbatively compute the renormalization scale running of scale-dependent quantities.

\subsubsection{Main characteristics of the setup}

We decide to perform three repetitions of the step scaling procedure, i.e. we start at a renormalization scale $\mu_0$ and then compute
succesively $\Sigma^X_{\Gamma}(\mu_0, \mu_1\!=\!2\mu_0)$, $\Sigma^X_{\Gamma}(\mu_1, \mu_2\!=\!2\mu_1)$ and $\Sigma^X_{\Gamma}(\mu_2, \mu_3\!=\!2\mu_2)$.
As it will turn out, these three values of the step scaling function will allow us to link non-perturbatively 
the scales ranging from around 1.5 GeV up to 17 GeV. The lower energy is accessible in a large volume 
simulation where the hadron matrix elements are evaluated, whereas the upper energy allows
for a safe translation to the infinite volume continuum $\MSb$ scheme. 

In order to stay on the line of constant physics, we fix the renormalization scale to be a fraction of the total simulated
spatial extent,
\begin{equation}
\mu L = \textrm{fixed} = \frac{L}{\sqrt{X_0^2}}.
\label{eq. scale}
\end{equation}
Since we always extrapolate our results to the chiral limit, $\mu$, $L$ and the the lattice spacing are the only relevant 
scales in our problem. Hence, the estimation of $\Sigma^X_{\Gamma}(\mu, 2\mu)$ necessitates two sets of lattices: one with 
spatial extent $L$ and the second with $2L$. 

\setlength{\tabcolsep}{5pt}
\begin{table}
\begin{center}
\begin{tabular}{|c| c c c c | c |}
\hline
step & 32/64 & 24/48 & 16/32 & 8/16 & volumes\\
\hline
1 & $\beta = 9.50(7)$ & ${\bf \beta = 9.00}$ & $\beta = 8.62(7)$ & $\beta = 7.90(13)$ & $L_1/\mathcal{L}_1$\\
2 & $\beta = 8.62(7)$ & $\beta = 8.24(6)$ & $\beta = 7.90(13)$ & $\beta = 7.18(2)$  & $L_2/\mathcal{L}_2$\\
3 & $\beta = 7.90(13)$ & $\beta = 7.56(11)$ & $\beta = 7.18(2)$ & $\beta = 6.61(2)$  & $L_3/\mathcal{L}_3$\\
\hline
\end{tabular}
\caption{\label{ensemble}Set of generated ensembles for each step of the step scaling technique. For each step of step scaling, the lattice volumes are matched to yield compatible effective couplings. The table header gives the lattice extents in the spatial direction. Each entry in the table contains the appropriate value of matching $\beta$ and its error, which is propagated to account for mismatching effects. $\beta=9.00$ given in bold is the starting point and hence has no associated uncertainty. \label{ensembles}}
\end{center}
\end{table}

Our starting point is an ensemble with $\hat{L}\equiv L/a=24$ at $\beta=9.0$ (with a corresponding, but unknown before the step scaling procedure, lattice spacing) which fixes our spatial extent $L$ in the first step of our procedure. 
We denote it by $L_1$, and the corresponding one of the $\hat{L}=48$ lattice by $\mathcal{L}_1 = 2 L_1$. 
We fix $\hat{L}$ to $8,\,16,\,24$ and $32$ and find the corresponding lattice spacing (or inverse bare coupling $\beta$) such that the spatial extent is always equal to $L_1$ (we describe the details of this procedure in the following section). We end up with a set of lattices as described by the first line of Tab.~
\ref{ensembles}. We have at our disposal four pairs of ensembles with equal physical volumes (and their doubles) and with different lattice spacings.
This defines our line of constant physics and allows to perfom the continuum extrapolation of the step scaling function.

We can now choose one particular renormalization scale given by $\mu_0 = 1/\sqrt{X^2}$ with $\hat{X}\equiv X/a=(1,1,1,1)$ at $\beta=7.90$ and 
evaluate the following ratios
\begin{align}
\frac{Z_{\Gamma}(\hat{X}=(1,1,1,1))}{Z_{\Gamma}(\hat{X}=(2,2,2,2))}\Big|_{\beta=7.90}, \qquad & \qquad \frac{Z_{\Gamma}(\hat{X}=(2,2,2,2))}{Z_{\Gamma}(\hat{X}=(4,4,4,4))}\Big|_{\beta=8.62},\nonumber  \\
\frac{Z_{\Gamma}(\hat{X}=(3,3,3,3))}{Z_{\Gamma}(\hat{X}=(6,6,6,6))}\Big|_{\beta=9.00}, \qquad & \qquad \frac{Z_{\Gamma}(\hat{X}=(4,4,4,4))}{Z_{\Gamma}(\hat{X}=(8,8,8,8))}\Big|_{\beta=9.50},
\end{align}
where the lattice spacings (or equivalenty $\beta$) were tuned in such a way that in terms of the physical distance, we can write in a symbolic way:
\begin{equation}
(1,1,1,1)\Big|_{\beta=7.90} \equiv 
(2,2,2,2)\Big|_{\beta=8.62} \equiv 
(3,3,3,3)\Big|_{\beta=9.00} \equiv 
(4,4,4,4)\Big|_{\beta=9.50}
\end{equation}
as well as
\begin{equation}
(2,2,2,2)\Big|_{\beta=7.90} \equiv 
(4,4,4,4)\Big|_{\beta=8.62} \equiv
(6,6,6,6)\Big|_{\beta=9.00} \equiv 
(8,8,8,8)\Big|_{\beta=9.50}.
\end{equation}
Let us stress again that by virtue of Eq.~\eqref{eq. scale}, the two estimates of the renormalization constant entering the above ratios
differ only by their renormalization scale, all other physical parameters being equal.

Having four estimates of the step scaling function $\Sigma^X_{\Gamma}(\mu, 2\mu,a)$ at different lattice spacings, allows us to 
extrapolate to the continuum and compute the step scaling function $\Sigma^X_{\Gamma}(\mu, 2\mu)$, Eq.~\eqref{eq. sigma}.

In the second step of the step scaling procedure, we work in a larger physical spatial extent $L_2 = 2 L_1$ and 
$\mathcal{L}_2$ = $2 \mathcal{L}_1$. Hence, we can reuse some of the ensembles from the previous step. For example, the $\hat{L}=16$ ensemble
at $\beta = 7.90$ used in the first step can be used again in the second step.
Similarly, in the third step of the procedure we can reuse two ensembles generated during the second step. Tab.~\ref{ensembles}
summarizes all the $\beta$ and $\hat{L}$ values needed for our study. For the coarsest ensemble, with the physical spatial extent $L_3 = 4 L_1$ and $\beta=6.61$, we can make contact with large volume simulations where the dimensionful lattice
spacing can be fixed in terms of the $r_0$ distance (see Sec.~\ref{sec:ensembles}).

In the discussion above, we demonstrated the concept of the step scaling method using a particular renormalization scale
given by $\mu_0 = 1/\sqrt{X^2}$ with $X/a=(1,1,1,1)$. However, one has also the possibility of computing the step scaling
function for other renormalization scales as well. In particular, we repeated our analysis for the renormalization scales given by
$\mu_0 = 1/\sqrt{X^2}$ with $X/a=(1,1,1,0)$ (four permutations averaged over\footnote{The cut-off effects are slightly different for the three cases where the zero coordinate is spatial and for the one where it is temporal. However, these differences are in practice much smaller than our statistical errors and hence we combine all four cases.}) and $\mu_0 = 1/\sqrt{X^2}$ with $X/a=(1,1,0,0)$ (six permutations with averaging). We excluded the data for the type of points $X/a=(1,0,0,0)$,
since they are known to be affected by very large cut-off effects. We label the three above possibilities by points of type IV, III and II, respectively.

\subsubsection{Translation to $\MSb$}
Since we want to compare finally with the running obtained in the $\MSb$ scheme, after checking that our volumes are large enough, we convert $\Sigma^X_{\Gamma}(\mu, 2\mu)$ to the latter, denoted by $\Sigma^{\MSb}_{\Gamma}(\bar\mu, 2\bar\mu)$, using 4-loop conversion formulae \cite{Chetyrkin:2010dx}.
Note that $\bar\mu=2e^{-\gamma_E}\mu\approx1.12\mu$, due to using a redefined scale in this reference. 
For more details on this step, we refer to our earlier work \cite{Cichy:2012is}, Section 2.4.
We note that in the following, we skip the superscript indicating the renormalization scheme, since it is always clear from the context which one is meant at a given stage.

\subsection{Matching of ensembles for step scaling}
As we discussed in the previous section, in order to perform three steps of the step scaling procedure, we need 
a set of gauge ensembles with a broad range of bare coupling values, $\beta\in[6.61,\,9.50]$ and lattice 
sizes, $L/a=8,\,12,\,16,\,24,\,32,\,48,\,64$. Moreover, ensembles with $L/a=8,\,16,\,24,\,32$ of the same step scaling step need to have matched physical volumes, such that one can relate the energy scale yielded by a given type of points of one ensemble to the energy scales for the other ones (i.e. that the scale $\mu=1/\sqrt{X^2}$ corresponds to the same value in physical units for e.g. points $\hat{X}=(1,1,1,1)$ ($\hat{L}=8$), $\hat{X}=(2,2,2,2)$ ($\hat{L}=16$), $\hat{X}=(3,3,3,3)$ ($\hat{L}=24$) and $\hat{X}=(4,4,4,4)$ ($\hat{L}=32$)).
Such matching for lattices of size $L_1/a_1$ and $L_2/a_2$ can be done in terms of an effective renormalized coupling\footnote{Note that we want to use this effective coupling only for the matching procedure here. In particular, we do not aim at using this coupling to compute the running of the renormalized strong coupling.}.
In this work, we define it using a certain ratio of Wilson loops proposed by Creutz \cite{Creutz:1980zw},
\begin{equation}
R_C(l,t)\equiv\frac{W(l,t)W(l+a,t-a)}{W(l,t-a)W(l+a,t)}, 
\end{equation}
where $W(l,t)$ is a Wilson-loop with spatial size $l$ and temporal size $t$.
Plugging $W(l,t)\propto e^{-\alpha t/l}$ (for $t\gg1$), we get
\begin{equation}
R_C(l,t)=e^{-\alpha \frac{a}{l} \frac{a}{l+a}}. 
\end{equation}
Hence, the effective coupling can be defined as
\begin{equation}
\alpha_{\rm eff}=-\frac{l}{a}\frac{l+a}{a}\ln R_C(l,t).  
\end{equation}
For $l$, we always use $l=L/4$.

If the thus defined renormalized couplings are the same (up to the precision of the calculation) on lattices with size  $\hat L_1=L_1/a_1$ and $\hat L_2=L_2/a_2$, this implies that $L_1\approx L_2$ (physical volumes are the same) and, equivalently, $a_2=(\hat L_1/\hat L_2)a_1$.

\begin{figure}[t!]
\begin{center}
\includegraphics[width=0.65\textwidth,angle=270]{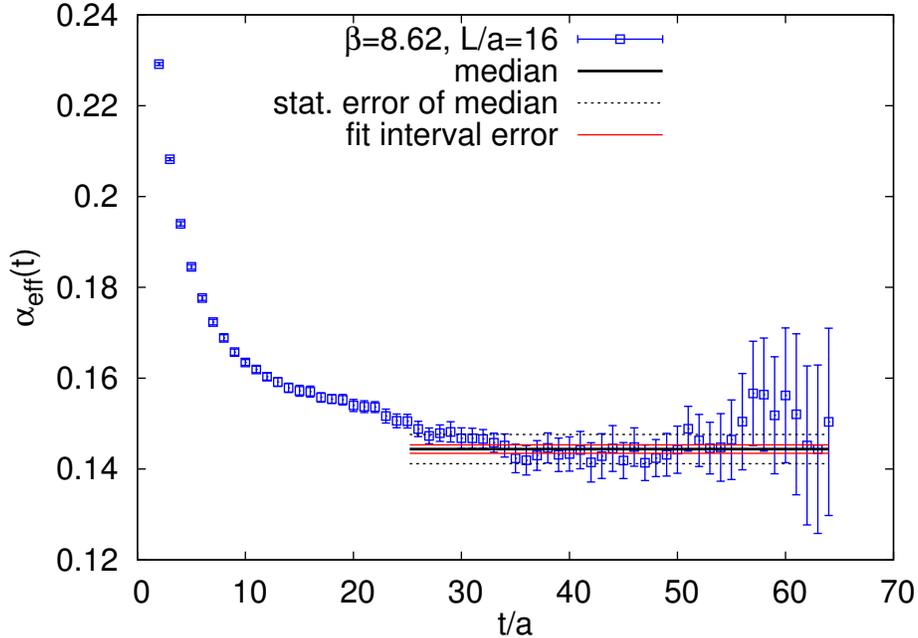}
\end{center}
\caption{\label{fig:eff} An example extraction of the effective coupling, for $\beta=8.62$ on a $16^3\times64$ lattice. The obtained result is $\alpha_{\rm eff}=0.1444(32)_{\rm stat}(9)_{\rm sys}$. The central value is given with the bold black line, the statistical error with the dashed black line and the systematic one by the thin solid red line. The error extraction procedure is explained in the text.}
\end{figure}

In Fig.~\ref{fig:eff}, we show an example extraction of the effective coupling, for one of our ensembles, at $\beta=8.62$ on a $16^3\times64$ lattice\footnote{Note that for reliable extraction of the effective coupling, we decided to have the temporal extent of lattices with $T=4L$ (for $L/a\leq32$) and $T=2L$ (for $L/a\geq48$).}.
The effective coupling reaches a plateau at large $t/a$, as expected.
We adopt a systematic procedure to extract a value from the plateau that takes into account the possibility of choosing different fitting ranges.
The procedure is analogous to the one described in Refs.~\cite{Cichy:2012vg,Banuls:2013jaa}.
In short, we consider all possible fit ranges $[t_{\rm min}/a,t_{\rm max}/a]$ with $t_{\rm min}/a=25,26,\ldots$ and $t_{\rm max}/a=39,40,\ldots$, i.e. each fit contains at least 15 consecutive data points.
Then, we build a weighted histogram of all fit results and define the systematic error as the average difference of the 16th/84th percentile with respect to the median (such that 68\% of the fit results are within one systematic error from the median).
The weights for each fit are given as $\exp(-\chi^2/{\rm dof})$, i.e.\ fits with larger values of $\chi^2/{\rm dof}$ are suppressed.
The smallest value of $t_{\rm min}/a$ is chosen in such a way that even the worst fits in terms of $\chi^2/{\rm dof}$ still contribute to the weighted histogram (i.e.\ have $\chi^2/{\rm dof}$ only slightly larger than 1, with best fits having $\chi^2/{\rm dof}\approx\mathcal{O}(0.1-0.2)$).
However, it is important to observe that the procedure is almost independent of the chosen smallest $t_{\rm min}/a$ -- fits that start before the actual pleateau has been reached are almost completely suppressed by the exponential weighting factor and hence only fits with $\chi^2/{\rm dof}\lesssim1$ have a significant contribution to the finally extracted value of the effective coupling.
Note that the number of fits that enter the histograms is of the order of a hundred or a few hundred (depending on the temporal extent of the lattice) and hence the histograms are to a good approximation Gaussian.
For each case, we repeat the whole procedure on 1000 bootstrap samples (with blocking) to obtain also the statistical error of the median and confirm that autocorrelations are under control (in all cases, we find $\tau_{\rm int}=\mathcal{O}(0.4-0.8)$ for the Wilson loop observables).
In the example shown in Fig.~\ref{fig:eff}, the dominant error is the statistical one.
However, in several cases, also the systematic error is important and thus our error estimation procedure is essential to obtain reliable results.

\begin{figure}[t!]
\begin{center}
\includegraphics[width=0.65\textwidth,angle=270]{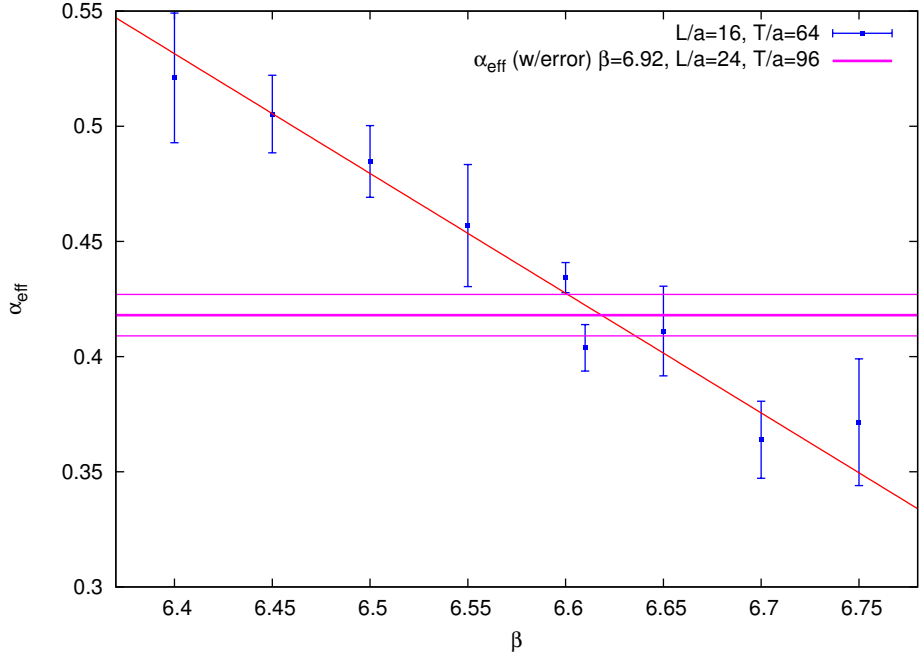}
\end{center}
\caption{\label{fig:match} Matching of a $16^3\times64$ lattice to yield the same effective coupling and physical volume as of our ensemble with $\beta=6.92$, $24^3\times96$. The matched value of $\beta$ is $\beta_{\rm match}=6.618(17)$, where the central value is given by the intersection of the thick magenta line (effective coupling of $\beta=6.92$, $24^3\times96$) and the red linear fit to the data points (effective couplings on $16^3\times64$ with varied $\beta$), while the error is, analogously, calculated from the intersection points with the thin magenta lines (reflecting the error of the effective coupling of $\beta=6.92$, $24^3\times96$).}
\end{figure}

Having established the procedure to extract the effective coupling, we can perform the matching of lattices. 
To show the reliability of our matching procedure, we show an example (Fig.~\ref{fig:match}) using an auxiliary ensemble (not used for evaluating the step scaling function) with inverse bare coupling $\beta=6.92$ and lattice size $24^3\times96$, for which we can cross-check the result by comparing to the prediction from Eq.~(2.6) in Ref.~\cite{Necco:2001xg}.
This formula expresses the Sommer parameter $r_0/a$ as a function of $\beta$ in the interval $\beta\in[5.7,\,6.92]$ and yields that $\beta=6.92$ with $L/a=24$ has the same volume as $\beta=6.613$ and $L/a=16$.
In our procedure, we want to find values of $\beta$ that yield the same effective coupling (and thus the same physical volume) on lattice sizes $L/a=16$.
We have generated several $16^3\times64$ ensembles with values of $\beta$ in the interval $\beta\in[6.4,\,6.75]$ and extracted the effective coupling for each ensemble.
Then, we have performed a linear fit (red solid line) to all data points and found the matching point as the intersection of the line from the fit with the line given by the effective coupling for the $\beta=6.92$, $24^3\times96$ ensemble.
The corresponding inverse bare coupling is $\beta=6.618$.
Considering the error of the effective coupling for $\beta=6.92$, we also determined the error of the matching value of $\beta$ to be 0.017, by observing intersections of the fit line with lines representing $\alpha_{\rm eff}\pm1\sigma$.
In Sec.~\ref{sec. systematic}, we discuss how to propagate this matching error to the level of renormalization constants, to account for effects of non-ideal matching.
Comparing the value $6.613$ from Eq.~(2.6) in Ref.~\cite{Necco:2001xg} with our estimate from the Creutz ratio effective coupling, $\beta=6.618(17)$, we conclude that our matching procedure is consistent with Ref.~\cite{Necco:2001xg}, well within the uncertainty that we find.
This gives us additional confidence in the adopted matching procedure.
Note also that the inverse bare coupling value $\beta=7.90$ appears in all three steps of the step scaling procedure, hence making the matching between different steps more robust.

Our next step is tuning the ensembles to maximal twist. We discuss it in the next subsection.

\subsection{Tuning to maximal twist}
Twisted mass lattice QCD enjoys the property of automatic $\mathcal{O}(a)$-impro\-ve\-ment when tuned to maximal twist \cite{Frezzotti:2003ni,Chiarappa:2006ae,Farchioni:2004ma,Farchioni:2004fs,Frezzotti:2005gi,
Jansen:2005kk}.
The commonly used criterion to achieve maximal twist is the vanishing of the PCAC quark mass, $m_{\rm PCAC}$.
In practice, it is enough if the criterion $m_{\rm PCAC}\leq0.1\mu_v$ holds \cite{Boucaud:2008xu}.

\begin{figure}[t!]
\begin{center}
\includegraphics[width=0.345\textwidth,angle=270]{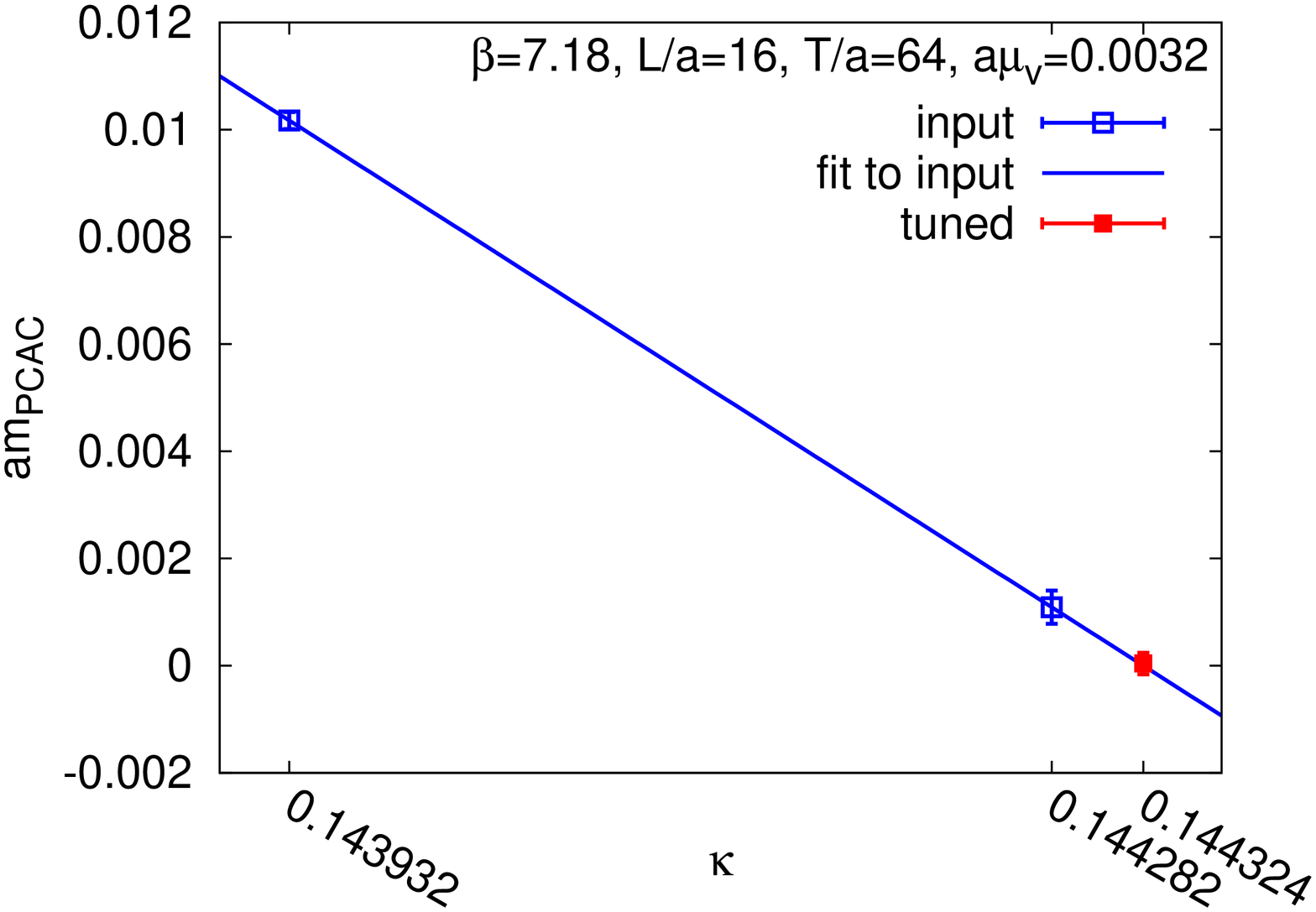}
\includegraphics[width=0.345\textwidth,angle=270]{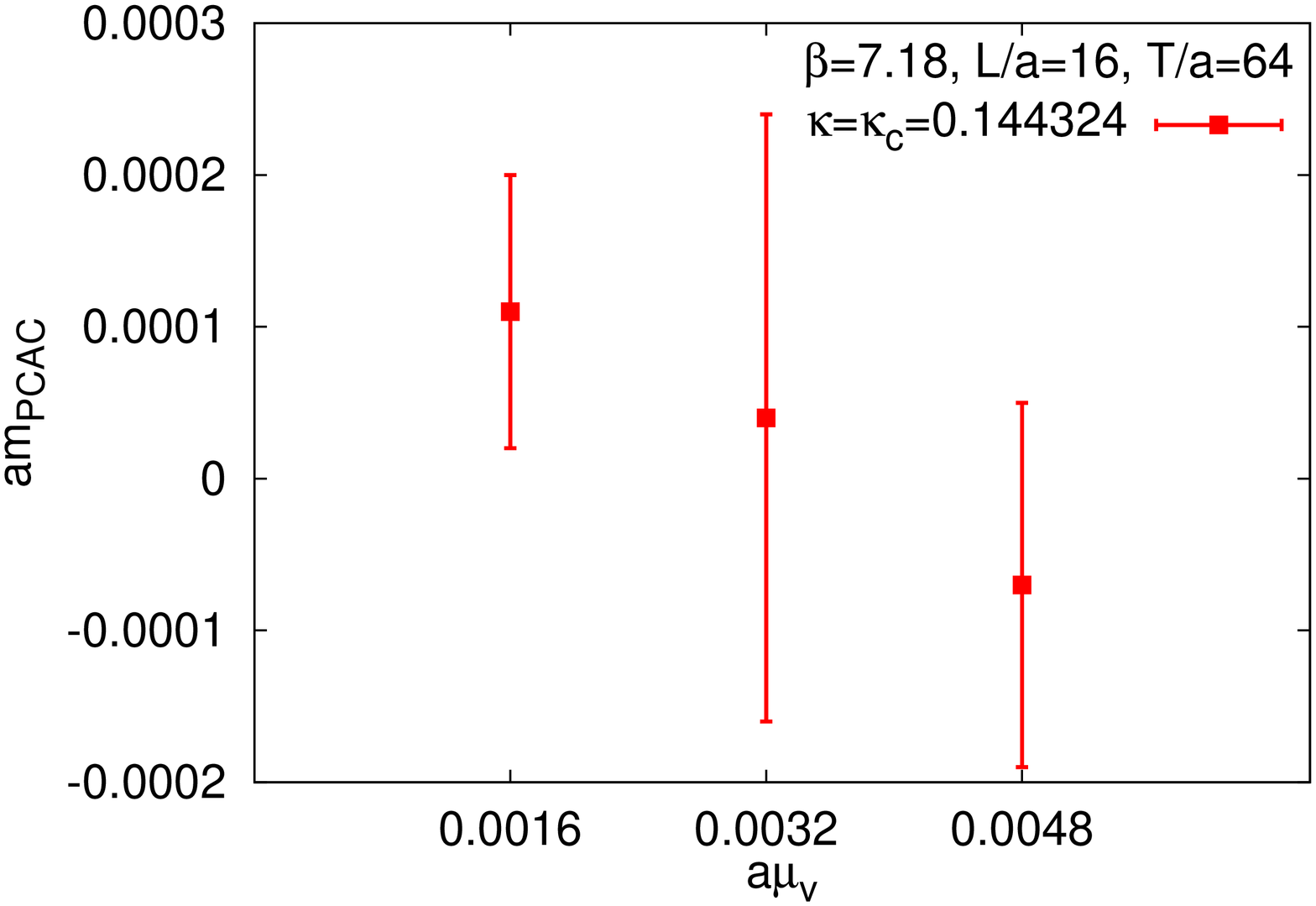}
\caption{\label{fig:PCAC}Illustration of tuning to maximal twist for the ensemble $\beta=7.18,\,L/a=16,\,T/a=64,\,a\mu_v=0.0032$. (left) Dependence of the PCAC mass on $\kappa$. Using the two input points (plotted with open blue squares, $\kappa=0.143932$ and $\kappa=0.144282$), we compute the derivative $\frac{\partial am_{\rm PCAC}}{\partial \kappa}$ (the slope of the fitted line) and predict the value of $\kappa_c=0.144324$ where the PCAC mass should vanish. The explict computation at this value of $\kappa$ (full red square) confirms that the PCAC mass vanishes and hence indeed this $\kappa=\kappa_c$. (right) Dependence of the PCAC mass at $\kappa=\kappa_c=0.144324$ on the valence quark mass $a\mu_v$. We take the same value of $\kappa_c$ for all masses.}
\end{center}
\end{figure}

For all our values of $\beta$ resulting from the matching procedure, we adopt the following strategy. We choose the smallest available volume such that $L/a\geq16$, we choose our intermediate valence quark mass, $a\mu_v=0.0032$ (in addition, we also use $a\mu_v=0.0016$ and $0.0048$, see below), and we guess a value of $\kappa_1$ for which the PCAC mass is small. Then, we compute the PCAC mass at another value of $\kappa=\kappa_2$, closer to its critical value, $\kappa_c$ (we know that the sign of the derivative $\frac{\partial am_{\rm PCAC}}{\partial \kappa}$ is negative).
This allows us to estimate the derivative $\frac{\partial am_{\rm PCAC}}{\partial \kappa}$ numerically, assuming linear dependence of $m_{\rm PCAC}$ on $\kappa$.
This, in turn, gives us an estimate of $\kappa_c$.
We illustrate this prescription graphically in Fig.~\ref{fig:PCAC} (left).
The PCAC masses for the two inital guesses, $\kappa_1$ and $\kappa_2$, are plotted with open blue squares.
They yield an estimate $\frac{\partial am_{\rm PCAC}}{\partial \kappa}=-25.9(1.4)$.
Since $am_{\rm PCAC}(\kappa_2)=0.00109(31)$, we obtain that $\kappa=0.144324$ should give a vanishing PCAC mass.
Indeed, an explicit computation at this value of $\kappa$ gives $am_{\rm PCAC}(\kappa=0.144324)=0.00004(20)$ (full red square in Fig.~\ref{fig:PCAC} (left)), a value compatible with the criterion $m_{\rm PCAC}\leq0.1\mu_v$.
For the other two valence quark masses, $a\mu_v=0.0016$ and $a\mu_v=0.0048$, we assume that the value of $\kappa_c$ can be taken as the one found for $a\mu_v=0.0032$.
The right panel of Fig.~\ref{fig:PCAC} illustrates that this is a reasonable procedure -- the dependence of $am_{\rm PCAC}$ on $a\mu_v$ is rather small (and approximately linear) for the considered changes in $a\mu_v$.

We take the value of $\kappa_c$ found in the above way also for other volumes at the same value of $\beta$. We monitor the value of $am_{\rm PCAC}$ for all ensembles and we observe that the condition $m_{\rm PCAC}\leq0.1\mu_v$ is always satisfied within statistical uncertainty.
Hence, we conclude that tuning to maximal twist is very robust for our ensembles and thus we can safely take our continuum limits assuming $\mathcal{O}(a^2)$ leading cut-off effects.

\subsection{Summary of generated ensembles}
\label{sec:ensembles}

Finally, we list in Tab.~\ref{tab:setup} all our generated ensembles for which computation of correlation functions is performed, i.e. we don't list the ones only used for matching.
For each ensemble, we use three values of the valence quark mass, $a\mu_v=0.0016$, $0.0032$ and $0.0048$, to perform the chiral limit extrapolation reliably.
The number of generated configurations is 200 for all ensembles, except for $L/a=8$ where we choose 1000.
We always use a step of 40 updates between saving configurations to reduce autocorrelations.
The autocorrelations are almost absent in Wilson loop observables ($\tau_{\rm int}=\mathcal{O}(0.4-0.8)$) and are under control in the correlation functions ($\tau_{\rm int}=\mathcal{O}(0.4-3)$).

As we mentioned above, at $\beta=6.61$ we can express the value of the lattice spacing in terms of the $r_0$ value.
Using the parametrization of Ref.~\cite{Necco:2001xg}, we find $r_0/a=12.76$ for this $\beta$, which together with our chosen value of $r_0$ in physical units (0.48(2) fm), yields 0.0376(16) fm for the lattice spacing.
This, together with our results for the matching of physical volumes, sets the scale for all our ensembles (see Tab.~\ref{tab:setup}).

\begin{table}[t!]
\begin{center}
\begin{tabular}{cccccccc}
\hline
$\beta$ & $a$ [fm] & $L/a$ & $T/a$ & $\kappa$ & nr of confs & step\\
\hline
9.50 & 0.00235(10) & 64 & 128 & 0.137032& 200 & 40\\
& & 32 & 128&  & 200 & 40\\
9.00 & 0.00314(13) & 48 & 96 & 0.138060& 200 & 40\\
& & 24 & 96 &  & 200 & 40\\
8.62 & 0.00470(20) & 64 & 128 & 0.138976& 200 & 40\\
& & 32 & 128&  & 200 & 40\\
& & 16 & 64&  & 200 & 40\\
8.24 & 0.00627(26) & 48 & 96 & 0.140016& 200 & 40\\
& & 24 & 96 &  & 200 & 40\\
7.90 & 0.00941(39) & 64 & 128 & 0.141173& 200 & 40\\
& & 32 & 128&  & 200 & 40\\
& & 16 & 64&  & 200 & 40\\
& & 8 & 32 &  & 1000 & 40\\
7.56 & 0.01254(52) & 48 & 96 & 0.142512& 200 & 40\\
& & 24 & 96&  & 200 & 40\\
7.18 & 0.01881(78) & 32 & 128 & 0.144324& 200 & 40\\
& & 16 & 64&  & 200 & 40\\
& & 8 & 32 &  & 1000 & 40 \\
6.61 & 0.03763(157) & 16 & 64 & 0.148162 & 200 & 40\\
& & 8 & 32 &  & 1000 & 40\\
\hline
\end{tabular}
\caption{\label{tab:setup}Complete list of ensembles used for step scaling. We give the inverse bare coupling $\beta$, the lattice spacing in fm (with its uncertainty given by our choice of $r_0=0.48(2)$ fm, the lattice size, the value of $\kappa$ that yields maximal twist, the number of generated (saved) configurations and the number of heatbath updates between saved configurations.}
\end{center}
\end{table}

\section{Estimating systematic uncertainties of the step scaling function}
\label{sec. systematic}

Lattice simulations are necessarily influenced by different kinds of systematic effects.
To compare to continuum, infinite-volume, massless perturbation theory, we need to reliably extrapolate to the continuum, infinite-volume and chiral limits, respectively.

\subsection{Chiral limit} 

As already mentioned above, the chiral limit extrapolation is performed using three valence quark masses. This is illustrated with a detailed example in Sec.~\ref{sec:example}.
In all cases, we observe behaviour compatible with a linear dependence on the valence quark mass. In several cases, actually, the dependence is mild enough such that a constant fit would also be appropriate.
In total, quark mass effects are well under control and they are taken into account by propagating the error from the linear extrapolation to the massless limit.

\begin{figure}[t!]
\begin{center}
\includegraphics[width=0.6\textwidth,angle=270]{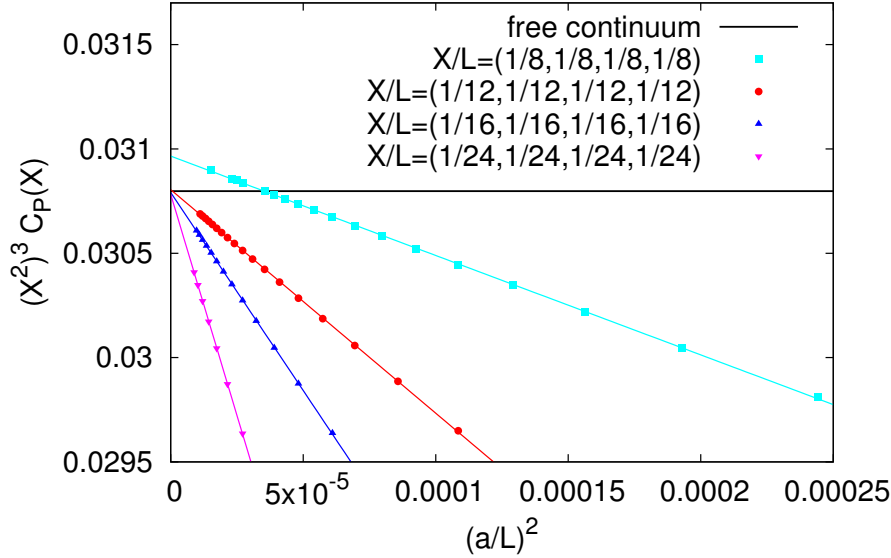}
\caption{\label{fig:freeFVE}Continuum limit scaling of the massless PP correlator $(X^2)^3 C_P(X)$ in the free theory. We look at points $(x/a,x/a,x/a,x/a)$ satisfying $x/L=1/4,\,1/8,\,1/12,\,1/16,\,1/24$ and change $L/a$, up to 336, to approach the finite-volume continuum limit. The case of $x/L=1/4$ is not shown as it yields a continuum limit more than 50\% higher than the infinite-volume one ($3/\pi^4$). For $x/L=1/8$, this deviation is around 0.5\% and for $x/L\leq1/12$ it is below 0.025\%. }
\end{center}
\end{figure}

\subsection{Finite volume effects} 
\begin{figure}[t!]
\begin{center}
\includegraphics[width=0.6\textwidth,angle=270]{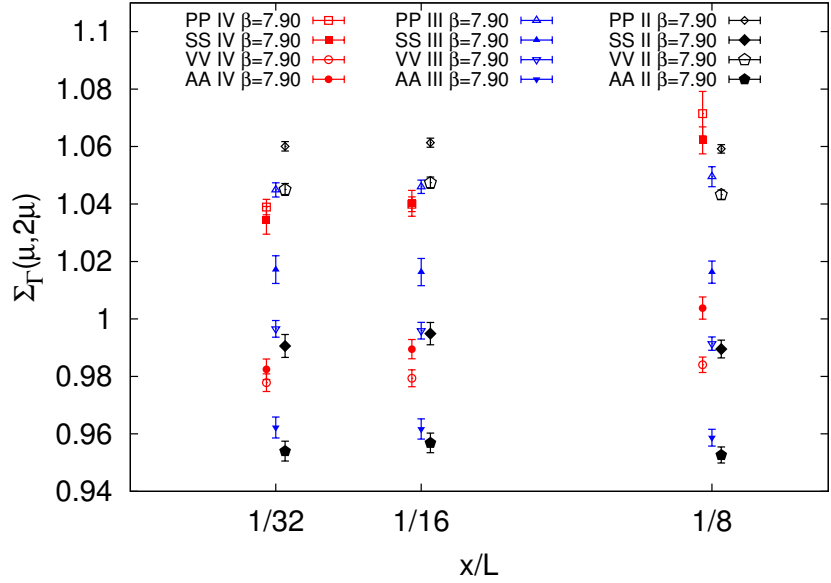}
\caption{\label{fig:FVE1}Finite volume effects in the step scaling function. We always take $x/a=1$ and $x/a=2$ and pairs of lattices with $L/a=8,\,16$, $L/a=16,\,32$ and $L/a=32,\,64$. In this way, $x/L$ changes from 1/8 to 1/16 and 1/32, respectively. Points of type II/IV are slightly shifted to the right/left for better visibility.}
\end{center}
\end{figure}

\begin{figure}[t!]
\begin{center}
\includegraphics[width=0.6\textwidth,angle=270]{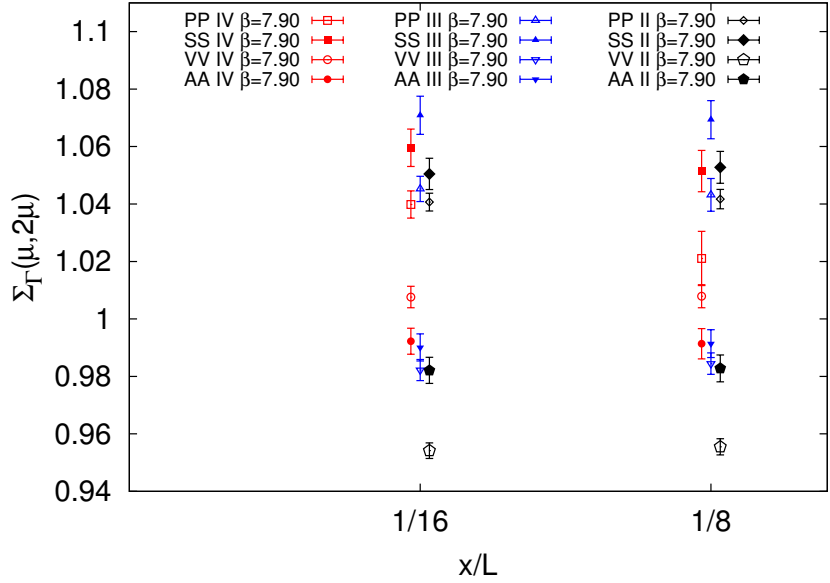}
\caption{\label{fig:FVE2}Finite volume effects in the step scaling function. We always take $x/a=2$ and $x/a=4$ and pairs of lattices with $L/a=16,\,32$ and $L/a=32,\,64$. In this way, $x/L$ changes from 1/8 to 1/16. Points of type II/IV are slightly shifted to the right/left for better visibility.}
\end{center}
\end{figure}

Another source of systematic uncertainties are finite volume effects (FVE).
To minimize them, we consider coordinate space points $(x/a,x/a,x/a,x/a)$, $(x/a,x/a,x/a,0)$ and $(x/a,x/a,0,0)$ (where $0$ is at different positions) that always satisfy the condition $x/L=1/8$ (for the temporal coordinate, we have $t/T=1/16$ or $t/T=1/32$).
This is the smallest value of $x/L$ for which FVE are expected to be small in the free theory, as shown in Fig.~\ref{fig:freeFVE}.
In case of larger values of $x/L$, we observe that the continuum limit in the free theory is different from the one of the infinite-volume free theory. 
With $x/L=1/4$, $x/L=1/8$ and $x/L=1/12$, the deviations of the continuum-extrapolated result from lattice data with respect to the free theory infinite-volume and continuum result are around 50\%, 0.5\% and 0.022\%, respectively.
Ideally, one should then choose $x/L<1/8$ in the interacting case.
However, the computational cost of taking e.g. $x/L=1/12$ is prohibitive, i.e.\ it would require generating lattices with $L/a=96$ for our finest lattice spacing of each step scaling step or using only three coarser lattice spacings and going to $L/a=72$.
Hence, it is essential to confirm that FVE are relatively small for $x/L=1/8$ that we choose for the interacting case.
We perform this check by investigating the dependence of the step scaling function $\Sigma_\Gamma(\mu,2\mu,a)$ at $\beta=7.90$, 
where we have four volumes available, with $L/a=8,\,16,\,32,\,64$. We look at $\Sigma_\Gamma(\mu,2\mu,a)$ 
calculated for points of type II, III and IV:
\begin{itemize}
\item with $x/a=1$ and $x/a=2$ and pairs of lattices with $L/a=8,\,16$, $L/a=16,\,32$ and $L/a=32,\,64$; this implies that $x/L$ changes from 1/8 to 1/16 and 1/32, respectively -- see Fig.~\ref{fig:FVE1},
\item with $x/a=2$ and $x/a=4$ and pairs of lattices with $L/a=16,\,32$ and $L/a=32,\,64$; thus $x/L$ changes from 1/8 to 1/16 -- see Fig.~\ref{fig:FVE2}.
\end{itemize}

Looking at these two figures, we arrive at very important conclusions. We confirm that FVE are smaller than statistical errors even with $x/L=1/8$, however \emph{only} when the pair of lattice extents $L/a=8,\,16$ is \emph{excluded}. For this pair, FVE can be statistically significant and up to 3\% (red squares and circles in Fig.~\ref{fig:FVE1} for points of type IV).
The pair $L/a=16,\,32$ is still fine from the point of view of FVE, even with $x/L=1/8$.
Hence, the ratio $x/L$ is not the only factor that determines the size of FVE -- in addition, the value of $(X/a)^2$ is also important and apparently FVE can be large if $(X/a)^2=4$, as happens on $L/a=8$ lattices.
The outcome of this test of FVE leads us to using the results from the pairs of lattices with $L/a=8,\,16$ in our continuum limit extrapolations only for checks of systematic effects from the fitting ansatz, while the central values are taken excluding such lattices.
Following this strategy, we can obtain good estimates of the infinite volume step scaling functions and hence apply infinite volume perturbation theory. 

\subsection{Continuum limit} 

We now move on to discuss the continuum limit extrapolations, which are the next source of systematic effects. 
We consider renormalization constants extracted from two types of correlation functions, ones to which tree-level correction has or has not been applied, see Sec.~\ref{sec. position space}.
We use four types of continuum extrapolation fitting functions:
\begin{equation}
\label{eq:A}
({\rm A})\quad\Sigma_\Gamma(\mu,2\mu,a)_{\rm corrected}= \Sigma_\Gamma(\mu,2\mu)_{\rm cont, A} + c_{\rm A} a^2,
\end{equation}
i.e.\ a linear fit in $a^2$, we take into account the three finest lattice spacings (for the fourth lattice spacing, we observe that deviations from the linear behaviour are usually significant) and only values of $\Sigma_\Gamma$ obtained with the tree-level correction (the non-corrected ones are not well described by a linear fitting ansatz due to larger discretization effects), in total three data points and two fitting parameters, $\Sigma_\Gamma(\mu,2\mu)_{\rm cont, A}$ and $c_{\rm A}$,
\begin{equation}
({\rm B})\quad\Sigma_\Gamma(\mu,2\mu,a)_{\rm corrected}= \Sigma_\Gamma(\mu,2\mu)_{\rm cont, B} + c_{\rm B} a^2 + d_{\rm B} a^4,
\end{equation}
which is a quadratic fit in $a^2$ to all four lattice spacings, again with only tree-level corrected ratios of renormalization constants, four data points and three fitting coefficients, $\Sigma_\Gamma(\mu,2\mu)_{\rm cont, B}$, $c_{\rm B}$ and $d_{\rm B}$,
\begin{equation}
({\rm C}1)\quad\Sigma_\Gamma(\mu,2\mu,a)_{\rm corrected}= \Sigma_\Gamma(\mu,2\mu)_{\rm cont, C} + c_{{\rm C}1} a^2,
\end{equation}
\begin{equation}
({\rm C}2)\quad\Sigma_\Gamma(\mu,2\mu,a)_{\rm non-corrected}= \Sigma_\Gamma(\mu,2\mu)_{\rm cont, C} + c_{{\rm C}2} a^2,
\end{equation}
i.e.\ a combined fit linear in $a^2$, using only three finest lattice spacings, in total six data points and three fitting parameters, $\Sigma_\Gamma(\mu,2\mu)_{\rm cont, C}$, $c_{{\rm C}1}$ and $c_{{\rm C}2}$,
\begin{equation}
({\rm D}1)\quad\Sigma_\Gamma(\mu,2\mu,a)_{\rm corrected}= \Sigma_\Gamma(\mu,2\mu)_{\rm cont, D} + c_{{\rm D}1} a^2 + d_{{\rm D}1}a^4,
\end{equation}
\begin{equation}
\label{eq:D2}
({\rm D}2)\quad\Sigma_\Gamma(\mu,2\mu,a)_{\rm non-corrected}= \Sigma_\Gamma(\mu,2\mu)_{\rm cont, D} + c_{{\rm D}2} a^2 + d_{{\rm D}2}a^4,
\end{equation}
which is a combined fit quadratic in $a^2$, using all four lattice spacings, in total eight data points and five fitting parameters, $\Sigma_\Gamma(\mu,2\mu)_{\rm cont, D}$, $c_{{\rm D}1}$, $c_{{\rm D}2}$, $d_{{\rm D}1}$ and $d_{{\rm D}2}$.

In case where no systematic effects are observed in continuum extrapolations, all four estimates of the continuum step scaling function should lead to the same results, i.e.\ $\Sigma_\Gamma(\mu,2\mu)_{\rm cont, A}=\Sigma_\Gamma(\mu,2\mu)_{\rm cont, B}=\Sigma_\Gamma(\mu,2\mu)_{\rm cont, C}=\Sigma_\Gamma(\mu,2\mu)_{\rm cont, D}\equiv\Sigma_\Gamma(\mu,2\mu)_{\rm cont}$.
Conversely, the differences between results from different fits provide an estimate of systematic uncertainties in the continuum extrapolation.

\subsection{Non-ideal matching} 

We also consider systematic effects related to non-ideal matching of physical volumes.
For given lattice volumes $L$ and $2L$ and a given value of $\beta$, we obtain the estimates of the lattice step scaling function.
Moreover, we know the matching uncertainty of $\beta$, denoted by $\Delta\beta$ and the values of the step scaling function obtained from ensembles at neighbouring value(s) of $\beta$, a smaller one, $\beta_-$ (for the first and second step scaling steps) and a larger one, $\beta_+$ (for the second and third step).
This allows us to estimate the derivative of the lattice step scaling function with respect to $\beta$:
\begin{equation}
\frac{\partial\Sigma_\Gamma(\mu,2\mu,a(\beta))}{\partial \beta}
\Big|_{\beta,\beta_-}\approx\frac{\Sigma_\Gamma(\mu,2\mu,a(\beta))- \Sigma_\Gamma(\mu,2\mu,a(\beta_-))}{\beta-\beta_-},
\end{equation}
\begin{equation}
\frac{\partial\Sigma_\Gamma(\mu,2\mu,a(\beta))}{\partial \beta}
\Big|_{\beta_+,\beta}\approx\frac{\Sigma_\Gamma(\mu,2\mu,a(\beta_+))- \Sigma_\Gamma(\mu,2\mu,a(\beta))}{\beta_+-\beta}.
\end{equation}
Our estimate of the matching uncertainty is then obtained as the product of $\frac{\partial\Sigma_\Gamma(\mu,2\mu,a)}{\partial \beta}
\Big|_{\beta,\beta_-}$ or $\frac{\partial\Sigma_\Gamma(\mu,2\mu,a)}{\partial \beta}
\Big|_{\beta_+,\beta}$ and $\Delta\beta$.
In the case of the second step of step scaling, we average over the two estimates of the uncertainty.

\subsection{Uncertainties of $\Lambda_{\MSb}$ and $r_0$} 

Our final uncertainties are related to the choice of the value of $\Lambda_{\MSb}$, which influences the conversion from the X-space scheme to the $\MSb$ scheme at a given scale $\mu$, and to the value of $r_0$ taken to set the scale. For the former, we take $\Lambda_{\MSb}^{(0)}=238(19)$ MeV \cite{Luscher:1993gh}.
For the latter, we choose $r_0=0.48(2)$ fm. Both uncertainties are propagated to the final results.

\section{Results}
\label{sec. results}

\subsection{Detailed example}
\label{sec:example}

We now illustrate our procedure, including our systematic error estimates, with a detailed example.
We choose the first step of step scaling and points of type IV, $(x/a,x/a,x/a,x/a)$.
We consider the scalar case, i.e.\ $\Gamma=\mathbbm{1}\equiv S$.

We aim to construct the step scaling function for the scale change from approx. 5.911 GeV  to 11.822 GeV\footnote{Note that the numbers below take into account the fact that $\bar\mu\approx1.12\mu=1.12/\sqrt{X^2}$.}, using the $\beta$ values:
\begin{equation}
\frac{Z_{S}(\sqrt{X^2}=0.0188\,\textrm{fm})}{Z_{S}(\sqrt{X^2}=0.0376\,\textrm{fm})} = 
\frac{Z_{S}(\bar\mu = 11.822\,\textrm{GeV})}{Z_{S}(\bar\mu = 5.911\,\textrm{GeV})}
\Big|_{\beta=7.90}, \
\Big|_{\beta=8.62}, \
\Big|_{\beta=9.00}, \
\Big|_{\beta=9.50}.
\end{equation}

\begin{figure}[p!]
\begin{center}
\includegraphics[width=0.33\textwidth,angle=270]{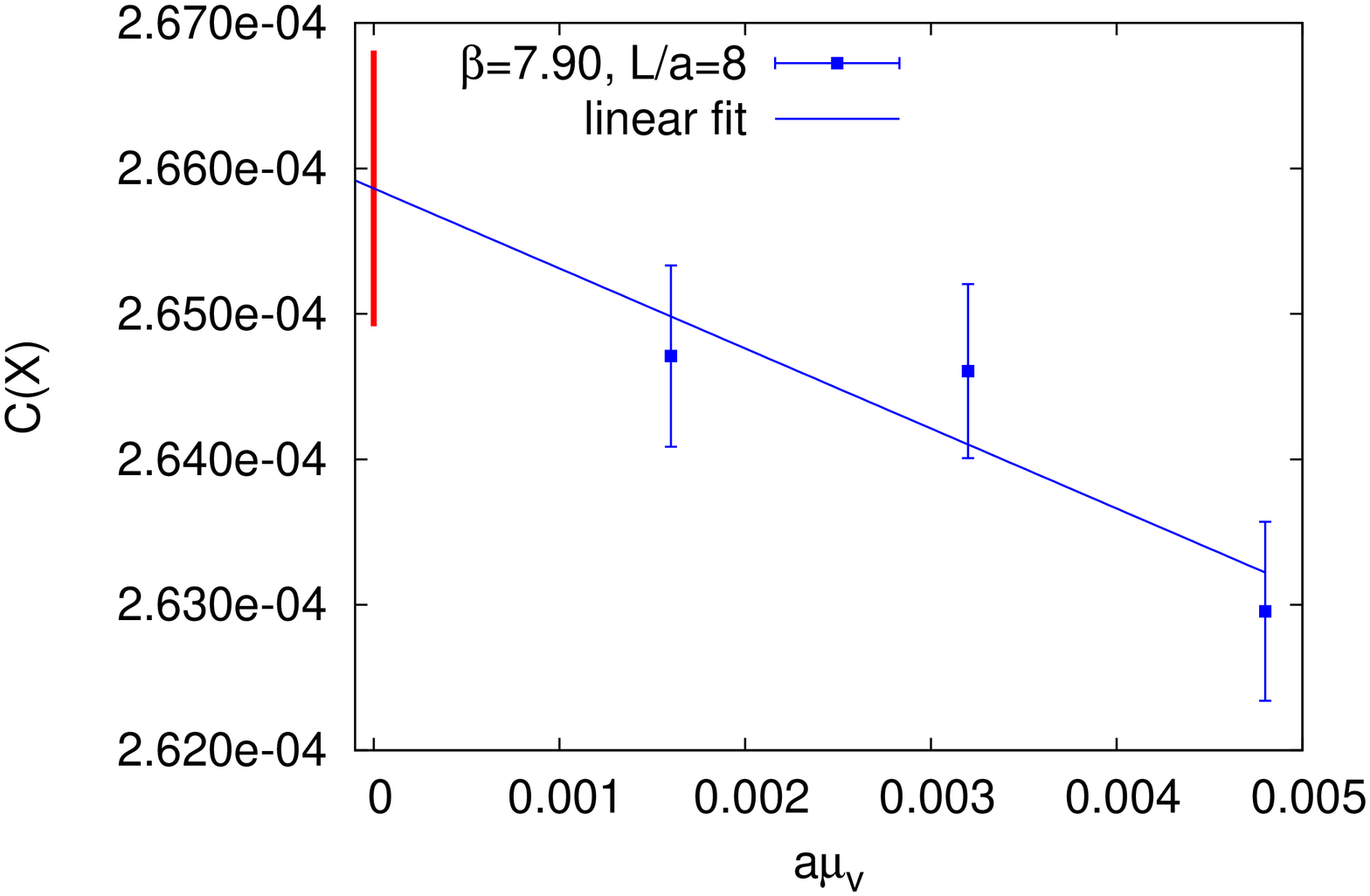}
\includegraphics[width=0.33\textwidth,angle=270]{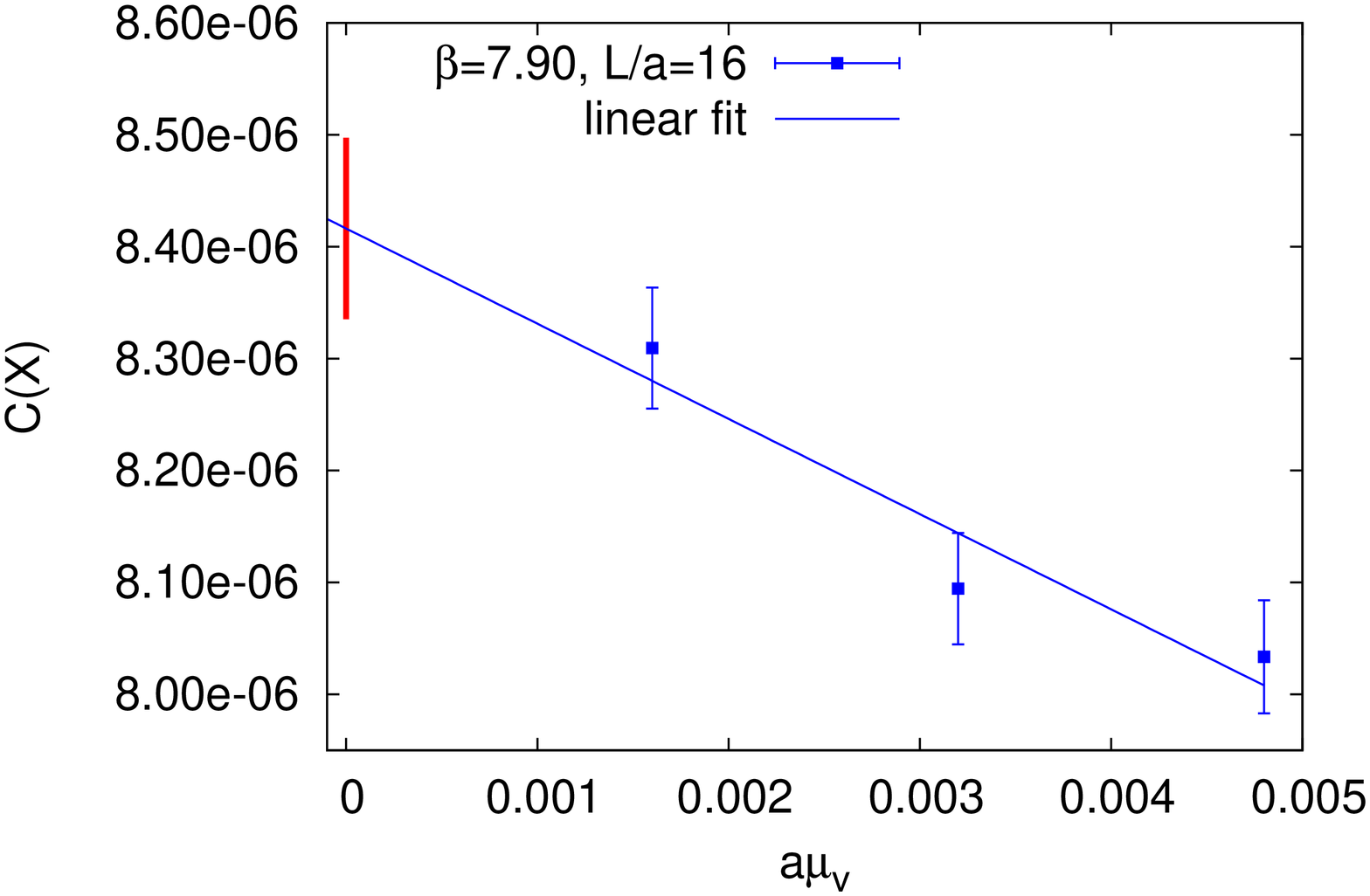}
\includegraphics[width=0.33\textwidth,angle=270]{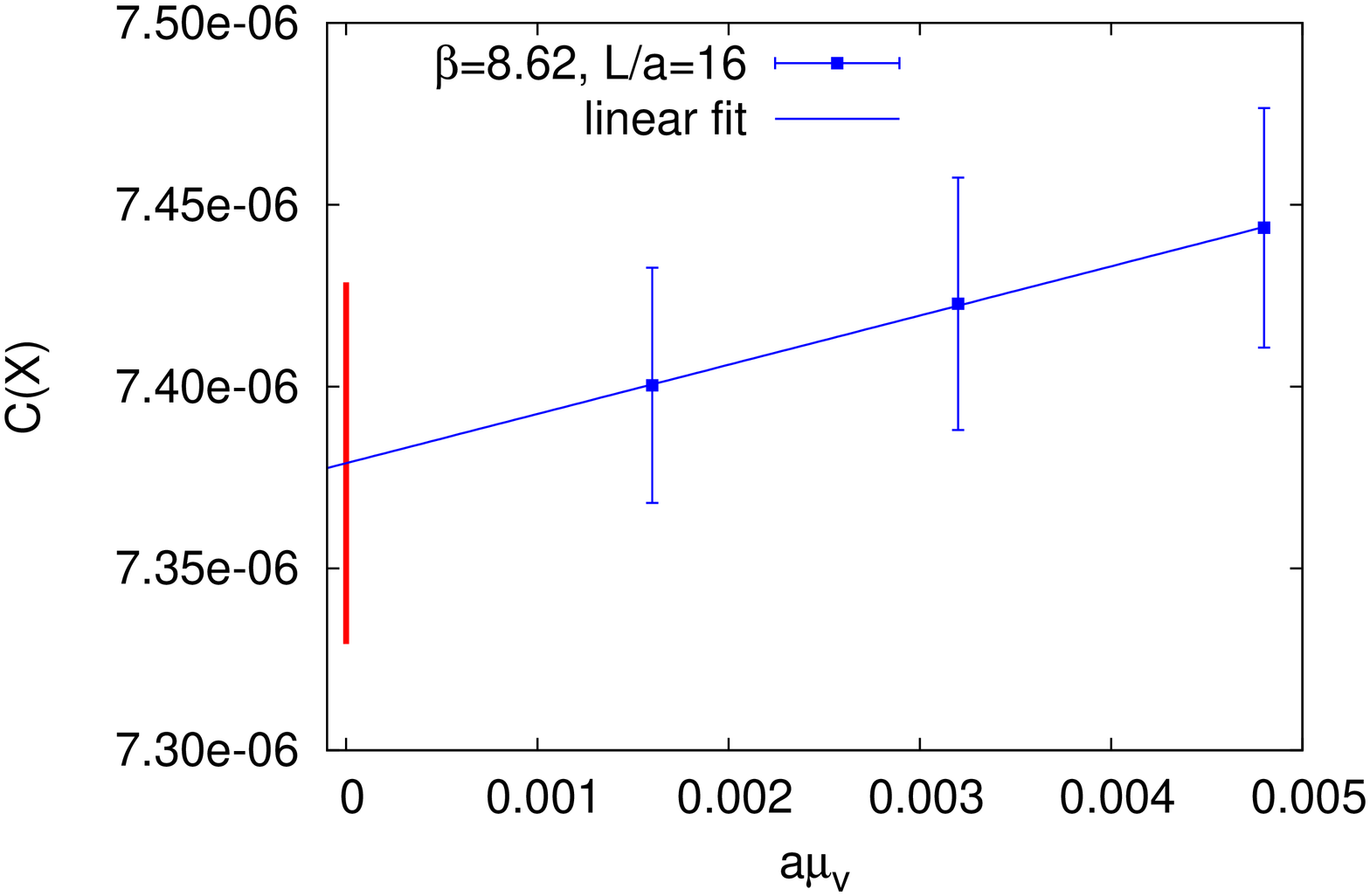}
\includegraphics[width=0.33\textwidth,angle=270]{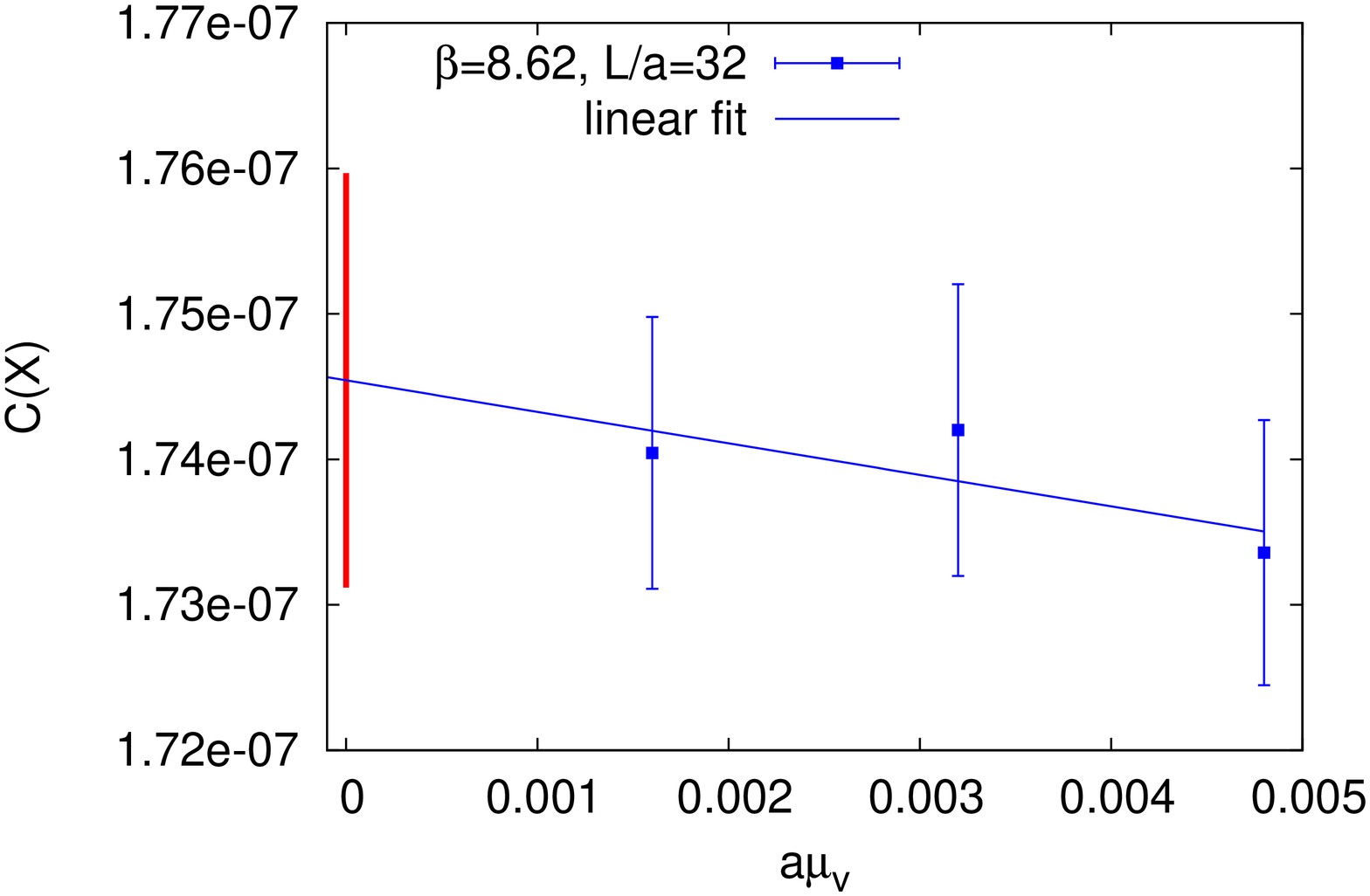}
\includegraphics[width=0.33\textwidth,angle=270]{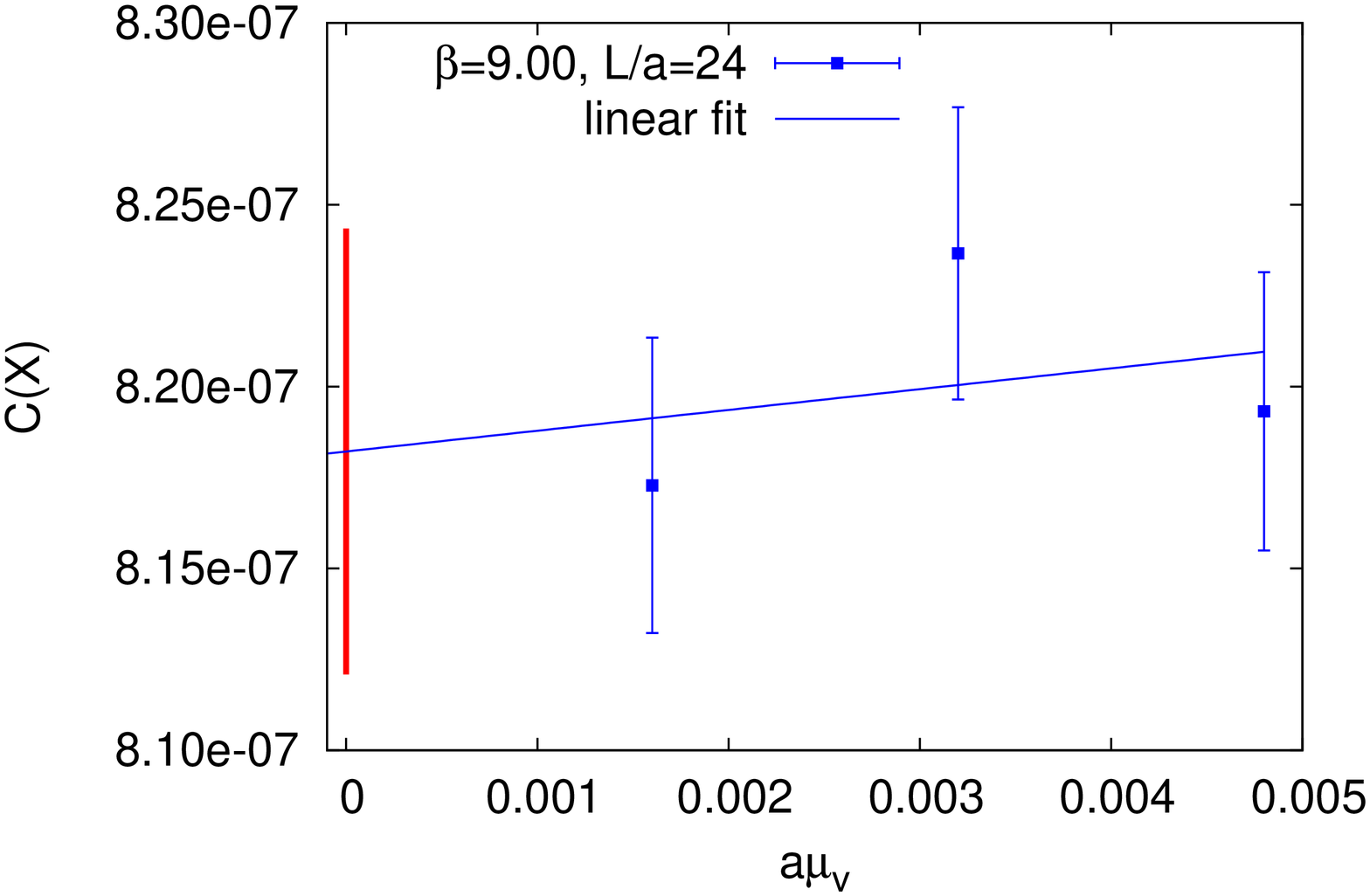}
\includegraphics[width=0.33\textwidth,angle=270]{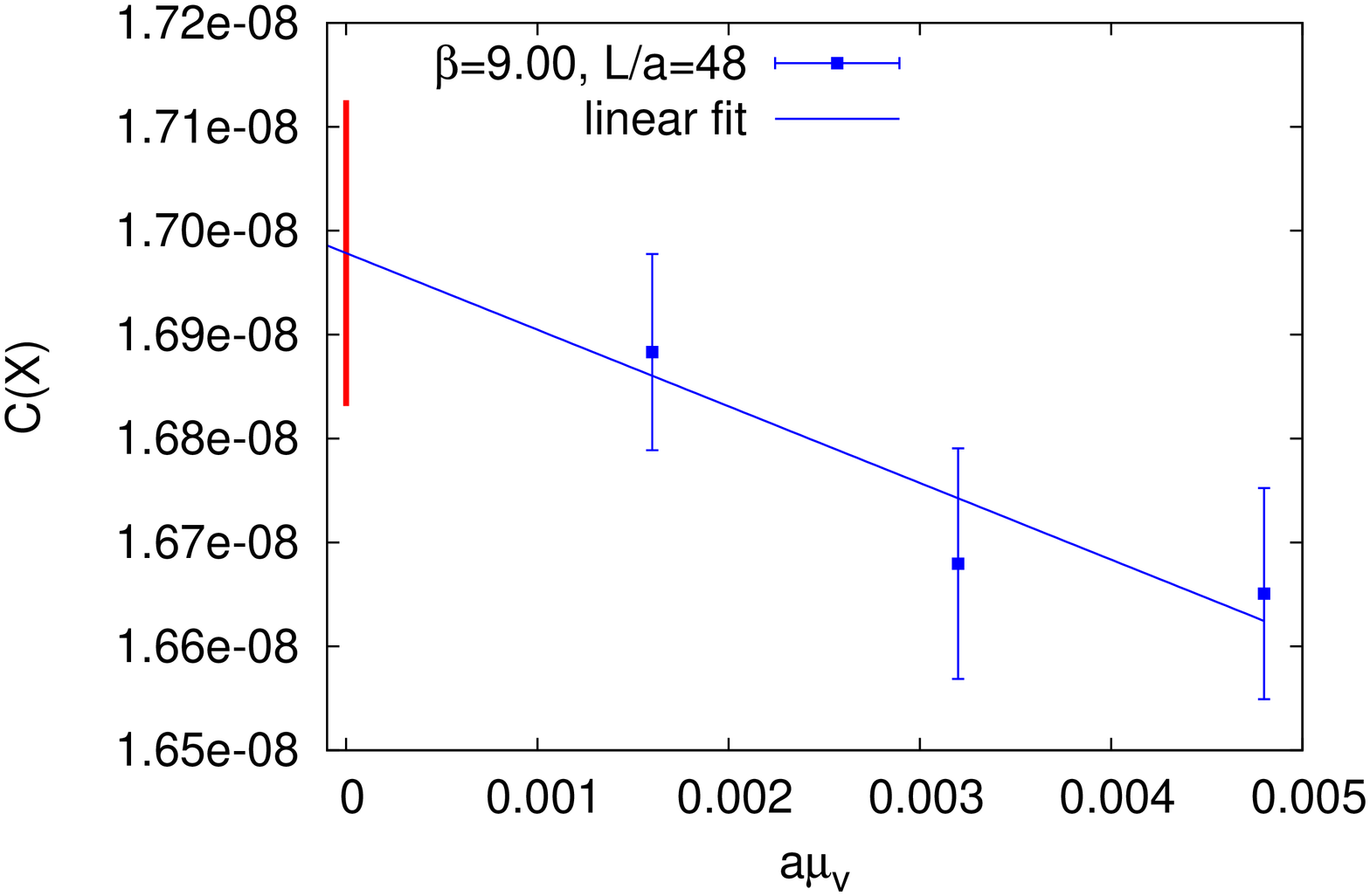}
\includegraphics[width=0.33\textwidth,angle=270]{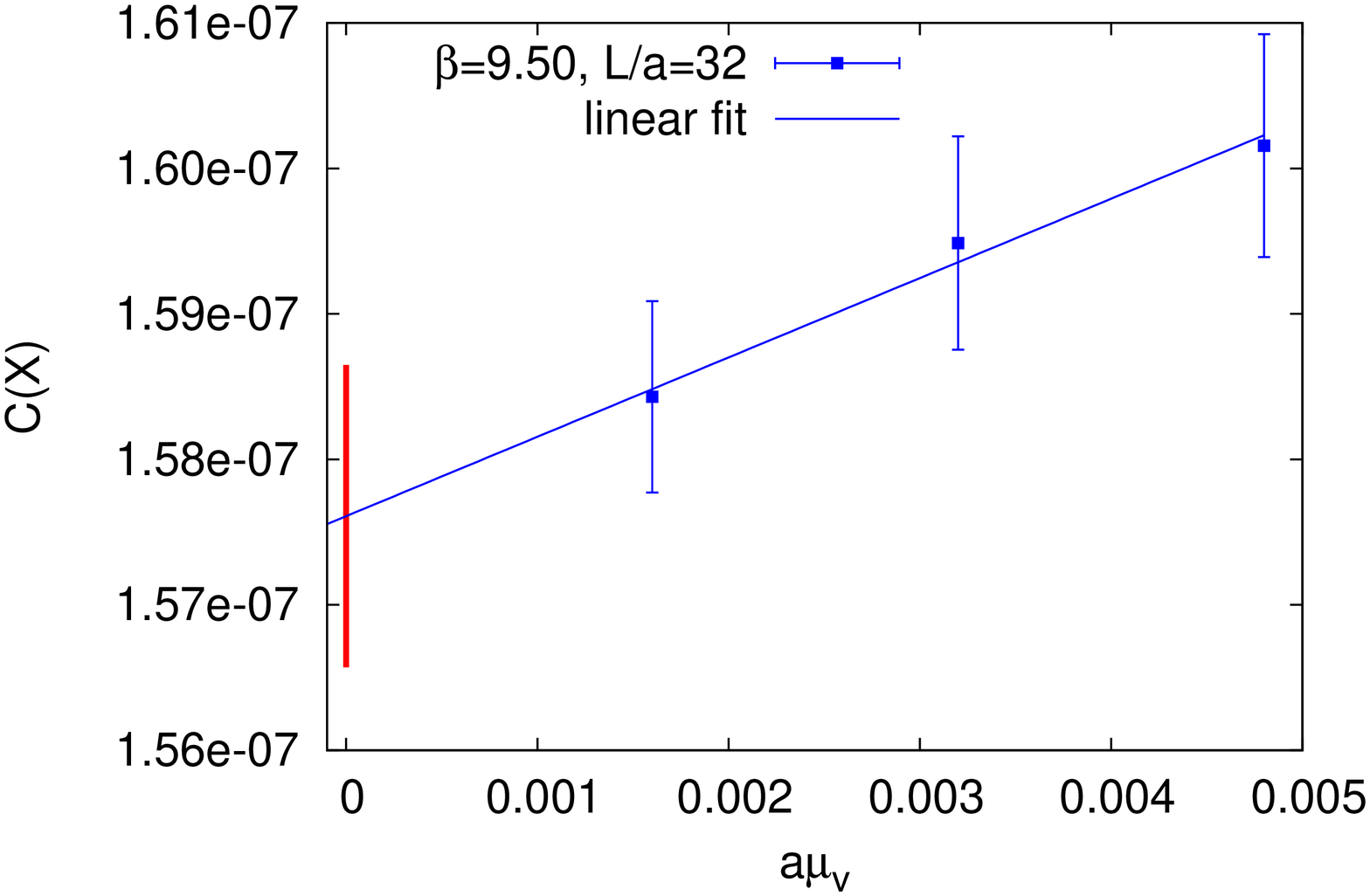}
\includegraphics[width=0.33\textwidth,angle=270]{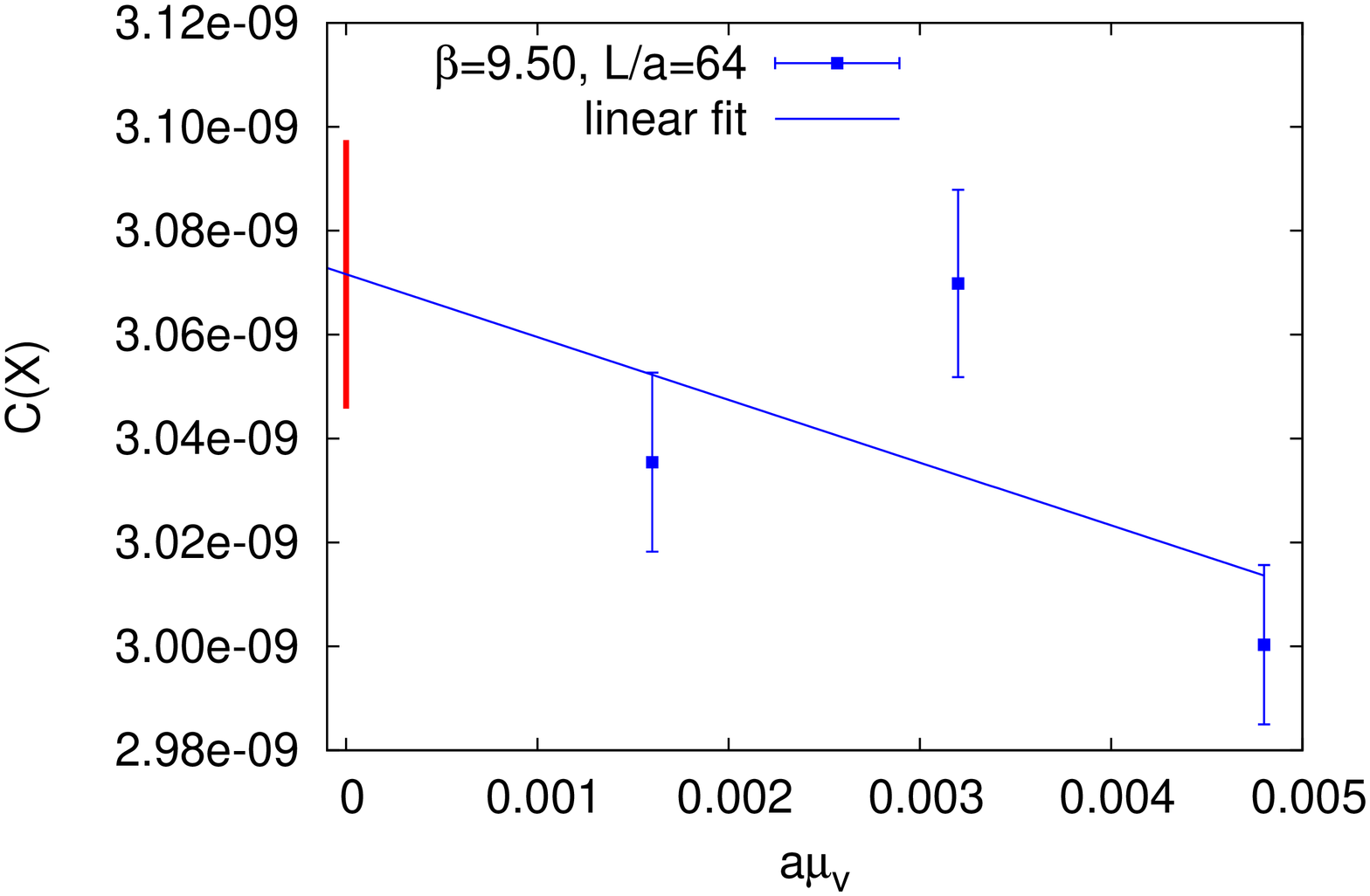}
\caption{\label{fig:chiral}Chiral extrapolations for the first step scaling step, scalar correlator, points of type IV, $(x/a,x/a,x/a,x/a)$, with $x/L=1/8$.}
\end{center}
\end{figure}

To construct the step scaling function at each lattice spacing, we first extrapolate the appropriate correlation functions to the chiral limit. This is illustrated in Fig.~\ref{fig:chiral}.
In all cases, the behaviour is consistent with our fitting ansatz, linear in the valence quark mass, $a\mu_v$.
We also note that the correlators at our lowest mass, $a\mu_v=0.0016$, are always compatible with the chiral limit extracted value for this case.
This confirms that quark mass effects are small.

\begin{figure}[t!]
\begin{center}
\includegraphics[width=0.6\textwidth,angle=270]{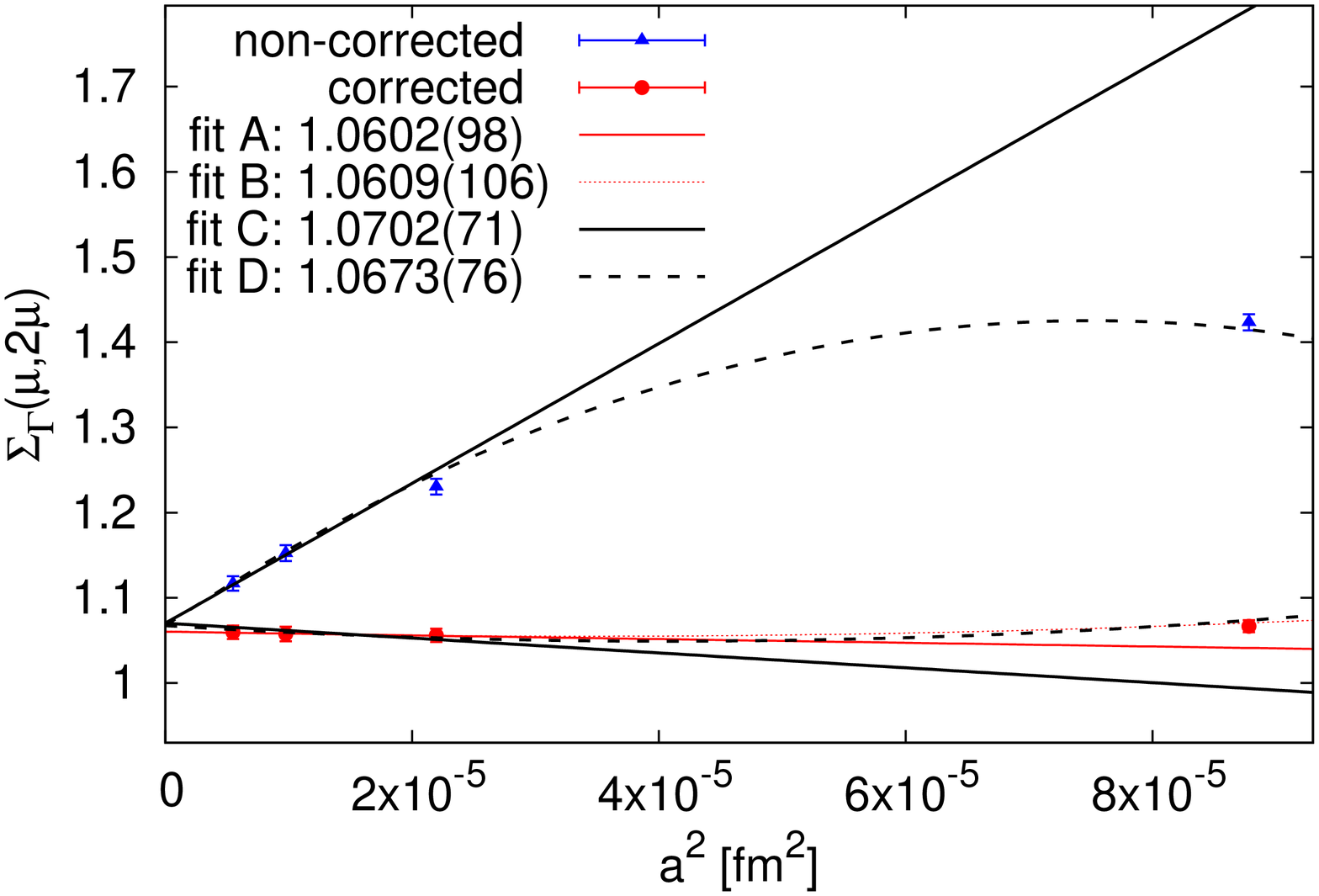}
\caption{\label{fig:cont}Different fitting ans\"atze for the continuum limit extrapolation. See Eqs.~(\ref{eq:A})-(\ref{eq:D2}) for the functional form. In the key, we give the values in the continuum limit.}
\end{center}
\end{figure}

The values of the correlation functions for appropriately chosen points are then used to construct the step scaling function $\Sigma_S(\mu,2\mu,a)$ at each of the four lattice spacings.
Our next step is to extrapolate to the continuum limit, using the four fitting ans\"atze defined in Eqs.~(\ref{eq:A})-(\ref{eq:D2}).
The extrapolations are shown in Fig.~\ref{fig:cont} and the continuum limit values, given in the key of the plot, agree well with one another within their errors.
As our preferred value, we take the one for fit C, since it yields the smallest statistical error.
The systematic uncertainty from the fitting ansatz is taken as the difference with respect to the continuum limit value for fit D, which includes one higher power of the lattice spacing and the data points from our coarsest lattice spacings.

Since we want to compare with continuum perturbation theory in the $\MSb$ scheme, we now convert our continuum limit value from fit C to this scheme.
The conversion factor is around 1.002 for these scales (for our third step of step scaling, involving scale change from around 1.5 GeV to 3  GeV, this factor can slightly exceed 1.01).
Hence, our final estimate of the step scaling function in the $\MSb$ scheme is:
\begin{equation}
\Sigma_S(5.911\,{\rm GeV},\,11.822 {\rm GeV})=1.0728(71)_{\rm stat}(30)_{\rm fit}(3)_{\rm match.}(2)_{\Lambda_{\MSb}^{(0)}}(9)_{r_0},
\end{equation}
where the errors are, respectively, statistical, one from the fitting ansatz uncertainty, one from the matching and the propagated uncertainty of $\Lambda_{\MSb}^{(0)}$ and $r_0$. Note that the $\Lambda_{\MSb}^{(0)}$ uncertainty enters only via the conversion formulae from the X-space scheme to the $\MSb$ scheme and is hence very small in the first step of step scaling (large energy scales).
The $r_0$ uncertainty enters via the conversion formulae and also via the absolute scale setting, thus influencing the estimate of the scale $\mu$.

The obtained value can be compared to the prediction of perturbation theory (PT) \cite{Chetyrkin:1997dh,Vermaseren:1997fq}:
\begin{equation}
\Sigma_S(5.911\,{\rm GeV},\,11.822 {\rm GeV})_{\rm PT}=1.0713(18),
\end{equation}
where the uncertainty comes from the uncertainty of $\Lambda_{\MSb}^{(0)}$.
We thus obtain good agreement between our computation and the prediction of continuum perturbation theory (with $N_f=0$).

\subsection{Full results for the step scaling function}
\label{sec:full}
In all other cases that we have computed, the analysis proceeds in a similar manner.
We summarize the procedure:
\begin{enumerate}
 \item Compute the relevant correlation functions in X-space, at three values of the valence quark mass.
 \item Extrapolate to the chiral limit.
 \item Apply the tree-level correction.
 \item Use the chirally extrapolated and tree-level corrected values of correlators to compute the step scaling function.
 \item Extrapolate to the continuum, using fit C or fit A (see below).
 \item Convert from the X-space to the $\MSb$ renormalization scheme.
 \item Calculate systematic uncertainties from the fitting ansatz (by comparing fits C/D or A/B, see below) and non-ideal matching and the uncertainties of $\Lambda_{\MSb}^{(0)}$ and $r_0$.
\end{enumerate}
Concerning points 5 and 7, we remark that combined fits (C and D) to tree-level corrected and non-corrected data points are possible \emph{only} when cut-off effects are under control for the non-corrected case. We find that this happens for the PP/SS correlators for points of type IV and III. For these correlators with points of type II, as well as for VV/AA correlators (all types of points), we note that the $\chi^2/\rm{d.o.f.}$ of the fits including non-corrected points indicate bad fits, most probably due to higher-order discretization effects in the non-corrected correlators. Thus, in such cases, we only consider fits A and B and take the central value from the former, while the latter serve to estimate the systematic uncertainty.

Our results for the continuum-extrapolated step scaling function, determined from both the scalar and the pseudoscalar correlator, are shown in Tab.~\ref{tab:step1} and the comparison with 4-loop $N_f=0$ continuum perturbation theory \cite{Chetyrkin:1997dh,Vermaseren:1997fq} is given in Tab.~\ref{tab:step1a}.
In most cases, we obtain good agreement between our lattice-extracted results extrapolated to the continuum limit and PT.
This concerns in particular the results from the scalar correlator, which are typically ($0-1.5$)-$\sigma$ away from PT.
However, we observe some regularities depending on the type of points that we consider -- points of type II(IV) tend to lie above(below) the PT result and points of type III tend to agree best with PT.
This suggests that cut-off effects are very different for these kinds of points, which is indeed plausible, taking into account that very different cut-off effects are observed even in the free theory.
A systematic treatment of these discretization effects (hypercubic artefacts) is highly desirable in order to obtain reliable results from the X-space scheme.

The results obtained from the pseudoscalar correlator are slightly worse in terms of agreement with PT, i.e. they tend to lie systematically below PT.
In most cases, they are still within $1.5$-$\sigma$, with the exception of the lowest scale $\mu$ (more than $2$-$\sigma$ too low).
We note, however, that much better agreement with PT is observed for points of type III and IV when \emph{not} using the tree-level correction for the step scaling function (for points of type II this makes the agreement even worse).
This strongly suggests that the cut-off effects for the PP case are much worse under control and, again, their better understanding is very important for the reliability of the X-space scheme.

\setlength{\tabcolsep}{2pt}
\begin{table}[t!]
\begin{center}
\begin{footnotesize}
\begin{tabular}{ccccc}
\hline
$\mu$ & $2\mu$ & point & $\Sigma_P^{\MSb}(\mu,2\mu)$  & $\Sigma_S^{\MSb}(\mu,2\mu)$ \\
${\rm [GeV]}$ & ${\rm [GeV]}$ & type & lattice & lattice \\
\hline
1.478 & 2.956 & IV &  1.0995(104)(66)(33)(13)(37) & 1.1134(121)(56)(37)(13)(37)\\ 
1.706 & 3.413 & III &  1.1027(91)(19)(36)(10)(29) & 1.1210(115)(6)(41)(11)(29)\\
2.090 & 4.180 & II &  1.1012(101)(33)(49)(8)(23) & 1.1337(140)(13)(52)(8)(23)\\
2.956 & 5.911 & IV &  1.0787(81)(31)(21)(4)(16) & 1.0856(90)(27)(21)(5)(16)\\
3.413 & 6.826 & III &  1.0743(72)(18)(19)(4)(14) & 1.0846(90)(14)(18)(4)(14)\\
4.180 & 8.360 & II &  1.0691(78)(22)(23)(3)(12) & 1.0961(109)(1)(22)(3)(12)\\
5.911 & 11.822 & IV &  1.0721(57)(22)(3)(2)(9) & 1.0728(71)(30)(3)(2)(9)\\
6.826 & 13.651 & III &  1.0736(57)(35)(4)(2)(9) & 1.0802(73)(8)(5)(2)(9)\\
8.360 & 16.719 & II &  1.0571(65)(24)(5)(2)(8) & 1.0755(91)(1)(7)(1)(8)\\
\hline
\end{tabular}
\end{footnotesize}
\caption{\label{tab:step1}Results for the step scaling function $\Sigma_{P/S}(\mu,2\mu)$, determined from the PP and SS correlators. We show the scale change, the type of points that were used  and our continuum-extrapolated result translated from the X-space to the $\MSb$ scheme. The uncertainties given are: statistical, one from the fitting ansatz, one from matching, from $\Lambda_{\MSb}^{(0)}$ and from $r_0$.}
\end{center}
\end{table}

\setlength{\tabcolsep}{4pt}
\begin{table}[t!]
\begin{center}
\begin{footnotesize}
\begin{tabular}{ccccc}
\hline
$\mu$ & $2\mu$ & point & $\Sigma_V^{\MSb}(\mu,2\mu)$  & $\Sigma_A^{\MSb}(\mu,2\mu)$ \\
${\rm [GeV]}$ & ${\rm [GeV]}$ & type & lattice & lattice \\
\hline
1.478 & 2.956 & IV & 0.9918(103)(1)(15)(2)(1) & 0.9931(86)(30)(9)(2)(1)\\
1.706 & 3.413 & III & 0.9968(108)(6)(22)(2)(1) & 0.9987(87)(20)(16)(2)(1)\\
2.090 & 4.180 & II & 1.0127(99)(15)(24)(1)(1) & 0.9683(69)(61)(23)(1)(1)\\
2.956 & 5.911 & IV &  1.0061(87)(10)(6)(1)(1) & 1.0039(70)(12)(6)(1)(1)\\
3.413 & 6.826 & III & 1.0122(92)(2)(10)(1)(0) & 1.0092(74)(19)(8)(1)(0)\\
4.180 & 8.360 & II & 1.0273(85)(22)(12)(1)(0) & 0.9848(59)(60)(8)(1)(0)\\
5.911 & 11.822 & IV & 1.0017(69)(12)(0)(1)(0) & 1.0009(58)(6)(2)(1)(0)\\
6.826 & 13.651 & III & 1.0103(77)(12)(2)(1)(0) & 1.0085(63)(27)(3)(1)(0)\\
8.360 & 16.719 & II &  1.0125(73)(14)(8)(1)(0) & 0.9823(52)(53)(1)(1)(0)\\
\hline
\end{tabular}
\end{footnotesize}
\caption{\label{tab:step2}Results for the step scaling function $\Sigma_{V/A}(\mu,2\mu)$, determined from the AA and VV correlators. We show the scale change, the type of points that were used and our continuum-extrapolated result translated from the X-space to the $\MSb$ scheme. The uncertainties given are: statistical, one from the fitting ansatz, one from matching, from $\Lambda_{\MSb}^{(0)}$ and from $r_0$.}
\end{center}
\end{table}

\setlength{\tabcolsep}{2pt}
\begin{table}[t!]
\begin{center}
\begin{footnotesize}
\begin{tabular}{cc|ccc|ccc}
\hline
 $\mu$ & $2\mu$ & $\Sigma_{P/S}^{\MSb}(\mu,2\mu)$ & $\Sigma_P^{\MSb}(\mu,2\mu)$ & $\Sigma_S^{\MSb}(\mu,2\mu)$ & $\Sigma_{V/A}^{\MSb}(\mu,2\mu)$ & $\Sigma_V^{\MSb}(\mu,2\mu)$ & $\Sigma_A^{\MSb}(\mu,2\mu)$\\
${\rm [GeV]}$ & ${\rm [GeV]}$ & 4-loop PT & lattice & lattice & exact & lattice & lattice\\
\hline
1.478 & 2.956 & 1.1318(68) & 1.0995(133) & 1.1134(144) & 1.0 & 0.9918(104) & 0.9931(92)\\
1.706 & 3.413 & 1.1206(56) & 1.1027(104) & 1.1210(126) & 1.0 & 0.9968(110) & 0.9987(91)\\
2.090 & 4.180 & 1.1080(44) & 1.1012(120) & 1.1337(152) & 1.0 & 1.0127(103) & 0.9683(95)\\
2.956 & 5.911 & 1.0919(31) & 1.0787(91) & 1.0856(98) & 1.0 & 1.0061(88) & 1.0039(71)\\
3.413 & 6.826 & 1.0866(27) & 1.0743(78) & 1.0846(94) & 1.0 & 1.0122(93) & 1.0092(77)\\
4.180 & 8.360 & 1.0802(23) & 1.0691(85) & 1.0961(112) & 1.0 & 1.0273(89) & 0.9848(85)\\
5.911 & 11.822 & 1.0713(18) & 1.0721(62) & 1.0728(78) & 1.0 & 1.0017(70) & 1.0009(58)\\
6.826 & 13.651 & 1.0682(16) & 1.0736(68) & 1.0802(74) & 1.0 & 1.0103(78) & 1.0085(69)\\
8.360 & 16.719 & 1.0643(15) & 1.0571(70) & 1.0755(92) & 1.0 & 1.0125(75) & 0.9823(74)\\
\hline
\end{tabular}
\end{footnotesize}
\caption{\label{tab:step1a}Comparison of lattice extracted results for the step scaling function $\Sigma_{P/S/V/A}(\mu,2\mu)$, determined from the PP, SS, VV and AA correlators, with continuum perturbation theory result (4-loop, $N_f=0$) in the (pseudo)scalar case and the exact continuum result of 1.0 in the (axial) vector case. The errors of the lattice result (see Tabs.~\ref{tab:step1}, \ref{tab:step2}) were combined in quadrature.}
\end{center}
\end{table}

We also computed the step scaling function $\Sigma_{V/A}(\mu,2\mu)$ for the vector and axial vector case (Tabs.~\ref{tab:step2}, \ref{tab:step1a}).
Obviously, its continuum value is always 1.0, but its lattice computation provides a cross-check of the method and of the cut-off effects for different types of points in coordinate space.
We observe that the expected value of 1.0 is well reproduced by our approach when using points of type III and IV.
The deviations from unity are at most 1.3-$\sigma$ and are of both signs, hence they are plausibly only statistical fluctuations.
However, for points of type II, there are more significant deviations, as large as 3-$\sigma$.
This suggests that cut-off effects in the vector and axial vector correlators are significant for the points of this type.

\begin{figure}[t!]
\begin{center}
\includegraphics[width=0.6\textwidth,angle=270]{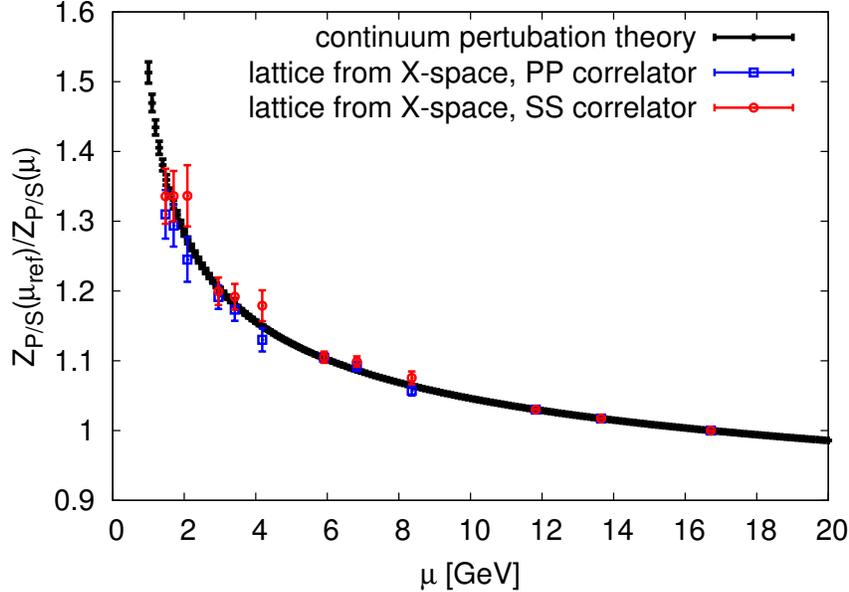}
\caption{\label{fig:run1}Running of the scalar/pseudoscalar renormalization constants. The black symbols correspond to continuum perturbation theory (with an uncertainty related to the uncertainty in the value of $\Lambda_{\MSb}^{(0)}$). The reference scale is $\mu_{\rm ref}=16.719$ GeV. The three rightmost points are the starting points for our step scaling procedure and hence have no errors. To the left of these, there are three groups of three points, corresponding to the three steps of step scaling and to the three types of points. Each rightmost point from a given group corresponds to points of type II, the middle one to type III and the leftmost to type IV.}
\end{center}
\end{figure}

\subsection{Comparison of the running with perturbation theory}
\begin{figure}[t!]
\begin{center}
\includegraphics[width=0.6\textwidth,angle=270]{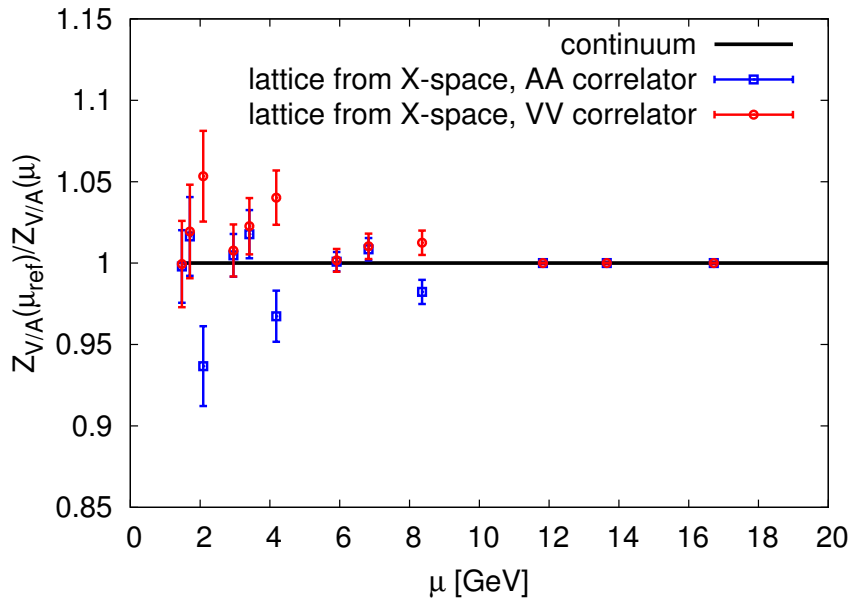}
\caption{\label{fig:run2}Running of the vector/axial vector renormalization constants. The black solid line corresponds to the continuum result of 1.0. The reference scale is $\mu_{\rm ref}=16.719$ GeV. The three rightmost points are the starting points for our step scaling procedure and hence have no errors. To the left of these, there are three groups of three points, corresponding to the three steps of step scaling and to the three types of points. Each rightmost point from a given group corresponds to points of type II, the middle one to type III and the leftmost to type IV.}
\end{center}
\end{figure}

As a summary of our results, we show two plots that illustrate the running of the renormalization constants, see Figs.~\ref{fig:run1} and \ref{fig:run2}.

The former shows the scalar/pseudoscalar case, i.e. the quark mass evolution with the energy scale.
We show all our step scaling steps and all types of points.
As already indicated, the running obtained from the scalar correlator is well reproduced, particularly for points of type IV and III (the left and the middle points in each group of three points corresponding to the same step).
Points of type II (rightmost points in the group of three) give a result that is approximately 1-$\sigma$ above the continuum curve.
When the running is extracted from the pseudoscalar correlator, we observe a possible tendency that the step scaling function is below its continuum value.
This is clearly visible for all types of points, although the result is always within 1-$\sigma$ of the continuum result even in the last step.
It is desirable to investigate more the source of this observation, which can still only be a statistical fluctuation.
We plan to address this issue in a forthcoming publication, where we will attempt to reduce hypercubic artefacts using strategies similar to ones applied already, with success, in momentum space, see e.g.\ Ref.~\cite{deSoto:2007ht}.

In Fig.~\ref{fig:run2}, we show the analogous ``running'' for the vector/axial vector case.
Obviously, the corresponding renormalization constants are scale-independent, hence the continuum value is exactly 1.0.
We obtain good agreement with this result for points of type III and IV.
However, for points of type II, deviations from unity are increasing when going to smaller energy scales, with 2-2.5-$\sigma$ discrepancy in the final step scaling step.
Interestingly, the behaviour is opposite for the vector and the axial vector case -- the former systematically overestimates the ratio and the latter underestimates it.
As for the (pseudo)scalar case, we plan to investigate this issue in more depth in our upcoming work.
Using techniques from Ref.~\cite{deSoto:2007ht} could help to reduce discretizations effects and bring also points of type II closer to unity.

\section{Discussion and outlook}
\label{sec. conclusions}
In this work, we investigated, for the first time, the step scaling technique using the coordinate space renormalization scheme.
This scheme has certain appealing properties with respect to e.g.\ the RI-MOM scheme, in particular it is gauge invariant and hence no gauge fixing is needed.
A feasibility analysis was performed in the quenched approximation, to reduce the computational cost to a tractable level.
We used the Creutz ratio to match lattices of different size such that they have the same physical volume and we performed three steps of step scaling to evaluate the running of the renormalization constants of the pseudoscalar and scalar densities, as well as of the vector and axial vector currents.
In principle, the computed running of the former, i.e. of the quark mass, allows for an evaluation of the renormalization group invariant renormalization constants \cite{Capitani:1998mq,Guagnelli:1999wp,DellaMorte:2005kg,Fritzsch:2012wq}, which we have not attempted here, since our aim was a feasibility study of the method.

We considered three types of points in position space and showed that they lead to somewhat different levels of consistency between the lattice-extracted and continuum-extrapolated results and continuum quenched perturbation theory.
The overall consistency with perturbation theory is satisfactory.
Only for the (axial) vector case with points of type II, we have encountered some difficulties, which motivates to explore other ways to reduce cut-off effects, which are different for different kinds of points in coordinate space.

We demonstrated that the X-space method can provide reliable results.
It will be interesting to investigate whether the remaining sources of systematic uncertainties, especially the hypercubic artefacts, can be controlled with even better precision than the one obtained here.
It would also be desirable to perform the analysis with points that are less likely to be influenced by residual finite volume effects, i.e. with a smaller ratio of the $x/y/z$ coordinates to the spatial lattice size $L$, e.g.\ of 1/12 instead of 1/8 that we used for this study.
This will, however, lead to significant increase of the computational cost, unless one uses only three lattice spacings, instead of four as in our current work.
Finally, it would be of interest to perform this study in the dynamical case.
Of course, the cost of such a step scaling procedure in the X-space scheme is substantial with dynamical fermions, but it provides a well-controlled computation of renormalization constants, allowing for a reliable non-perturbative renormalization.

\section*{Acknowledgements}
We thank Pan Kessel for his code to extract the Creutz ratios and discussions at the initial stage of this project. K.C.\ was supported in part by the Deutsche Forschungsgemeinschaft (DFG), project nr. CI 236/1-1 (Sachbeihilfe). Numerical simulations were carried out at the PAX cluster at DESY Zeuthen.

\bibliographystyle{jhep}
\bibliography{lattice}

\end{document}